\documentstyle[amssymb,aps,preprint]{revtex}

\begin{document}
\title{Derivative expansion of quadratic operators\\
in a general 't Hooft gauge}
\author{{\bf Vasilios Zarikas}}
\address{Nuclear and Particle Physics Division, Department of Physics, \\
University of Athens, GR - 157 71}
\maketitle

\begin{abstract}
A derivative expansion technique is developed to compute functional
determinants of quadratic operators, non diagonal in spacetime indices. This
kind of operators arise in general 't Hooft gauge fixed Lagrangians.
Elaborate applications of the developed derivative expansion are presented.
\end{abstract}

\pacs{Pacs numbers: 11.15.-q, 11.10.Wx, 02.30.Mv, 11.27.+d}

\section{Introduction}

A very powerful mathematical technique, that is used in many areas of
theoretical physics, is the heat kernel expansion of a quadratic operator.
It became very significant because it can help us to calculate quantum
corrections in quantum field theories and in quantum gravity.

The transition amplitudes are calculated in terms of a perturbation
expansion. If the path integral has a saddle point then we can perform a
functional Taylor expansion around it and keep the terms up to second order.
Saddle points are the solutions of the classical field equations which
dominate the path integral. In this way we find the one loop corrections as
a function of the determinant of the quadratic fluctuations of the quantum
fields.

The derivative expansion presented here extends the work of I. Moss, D. Toms
and S. Poletti \cite{moss}, \cite{moss1}. Their expansion is a modification
of the heat kernel asymptotic expansion. They expand in powers of covariant
derivatives.

The derivative expansion method is very useful when we want to evaluate one
loop corrections, in cases where the background field is not constant. This
is indeed the situation for the soliton solutions, as well for the tunneling
rates where the instanton solution is a function of the Euclidean space
coordinates. Recently theories that predict non topological soliton
solutions have attracted the interest. Similar expansions need to be
performed in order to study the fate of these non trivial time independent
classical solutions when quantum corrections are considered \cite{stewart},%
\cite{farhi},\cite{bagger},\cite{borelli}.

A more general version of the derivative expansion method is developed. This
can deal with an extended number of operators, non diagonal in spacetime
indices, covering all the interesting ones in gauge field theories. The new
formalism can deal with all the gauge choices, in 't Hooft gauge fixing. It
is thus very useful when we want to study the gauge dependency of the
various results.

\section{The derivative expansion}

Heat kernel expansions and the similar derivative expansions are useful when
the background field or fields are not homogeneous. In the conventional
estimations of the first quantum corrections, for example in the one loop
effective action, we assume that the classical scalar field (the vacuum
expectation value) is constant. In the case that this is not true,
derivatives of the background field do not vanish resulting in the
appearance of extra kinetic terms in the Lagrangian. In this case we have to
deal with second order operators of the form 
\begin{equation}
\Delta =-{\cal D}^{2}+X
\end{equation}
where ${\cal D}$ is the covariant derivative associated with the group
symmetry and $X$ a matrix in the same space with the group structure.

The heat kernel $K({\bf x,x}^{\prime },t),$ (${\bf x}$ represents a
spacetime point and $t$ a positive parameter) is a quantity that helps to
deal with the eigenvalues of the operator. If $u_{n}({\bf x})$ are the
normalized eigenfunctions of the operator $\Delta ,$ with corresponding
eigenvalues $\lambda _{n}\,,$ the heat kernel can be expressed as 
\begin{equation}
K({\bf x,x}^{\prime },t)=\sum_{n}u_{n}({\bf x})^{\dagger }u_{n}({\bf x}%
^{\prime })e^{-\lambda _{n}\,t}  \label{kernel}
\end{equation}
It is related with another useful quantity, the generalized $\zeta $
function 
\begin{equation}
\zeta ({\bf x,}s)=\sum_{n}\lambda _{n}^{-s}
\end{equation}
in the following way : 
\begin{equation}
\zeta ({\bf x,}s)=\frac{1}{\Gamma (s)}\int_{0}^{\infty }dt\,t^{s-1}{\rm tr}%
[K({\bf x,x,}t)]
\end{equation}
for $s>d/2,$ (with space dimensions $d$ ). It can also continued
analytically at $s=0.$ The above defined functions $K({\bf x,x}^{\prime },t)$%
, $\zeta ({\bf x,}s)$ are useful in computing one loop quantum corrections
since the functional determinant of the relevant quadratic operator can be
expressed as: 
\begin{equation}
\det \Delta =\prod\limits_{n}\lambda _{n}=\exp \left[ -\int d^{4}x\,\,\zeta
^{\prime }({\bf x,}0)\right]  \label{det}
\end{equation}
with 
\begin{equation}
\zeta ^{\prime }({\bf x,}0)=\frac{d\zeta ({\bf x,}s)}{ds}\mid _{s=0}
\end{equation}

The derivative expansion I closely follow, suggested in \cite{moss1}, is
based on a different expansion from that in Eq. \ref{kernel}. The expansion
can be written in powers of covariant or ordinary derivatives 
\begin{equation}
{\rm tr}[K({\bf x,x,}t)]=(4\pi t)^{-d/2}\sum_{i}A_{i}({\bf x,}t{\bf )}
\label{A}
\end{equation}
where $A_{i}({\bf x,}t{\bf )}$ is a function carrying $i$ derivatives of $X.$
The derivative expansion is a local expansion which sums contributions of a
given number of derivatives to all orders contrary to the heat kernel
expansion which sums different number of derivatives order by order and
therefore has, in some cases, worst infrared problems. These functions $%
A_{i}({\bf x,}t{\bf )}$ can be computed using the heat kernel equation
expanded in the momentum space. The heat kernel equation 
\begin{equation}
\Delta K=-\frac{\partial K}{\partial t}=-\stackrel{.}{K}
\end{equation}
can be written as 
\begin{equation}
\int d^{4}x\,\,\left[ k^{2}+X(k,{\bf x})\right] \,K({\bf x,x}^{\prime
},t)\,e^{-ik({\bf x-x}^{\prime })}=-\stackrel{\cdot }{K}(k,{\bf x}^{\prime
},t)
\end{equation}
where $K(k,{\bf x}^{\prime },t)$ is the following transformation of $K({\bf %
x,x}^{\prime },t)$ , 
\begin{equation}
K(k,{\bf x}^{\prime },t)=\int d^{4}x\,\,e^{-ik({\bf x-x}^{\prime })}\,K({\bf %
x,x}^{\prime },t)\;.  \label{trans}
\end{equation}
We also expand in powers of usual derivatives 
\begin{equation}
K(k,{\bf x},t)=K_{0}(k,{\bf x},t)\sum_{n}a_{n}(k,{\bf x},t)\;,\;a_{o}=1,%
\qquad X=\sum_{n}X_{n}\;,
\end{equation}
where $X_{n}$ is the part of the quadratic terms carrying $nth$ order
derivatives or in other words the $nth$ order kinetic terms while $X_{0}$
contains the mass terms. $K_{0}$ is the part of the heat kernel with no
derivatives and $a_{n}$ is the part of the heat kernel that carries $nth$
order derivatives divided by $K_{0}.$

We can now express the functions $A_{n}(x,t)$ in terms of the new defined
functions $a_{n}$. From expressions Eq. \ref{A}, Eq. \ref{trans} we get 
\[
\sum_{i}A_{i}({\bf x,}t{\bf )=}(4\pi t)^{d/2}\,{\rm tr}\int d^{4}x\int d\mu
(k)\,K(k,{\bf x}^{\prime },t)\,e^{ik({\bf x-x}^{\prime })}\delta ({\bf x-x}%
^{\prime })\Rightarrow 
\]
\begin{equation}
A_{n}({\bf x},t)=(4\pi t)^{d/2}\int d\mu (k)\,{\rm tr}[K_{0}(k,{\bf x,}%
t)\,a_{n}(k,{\bf x},t)]  \label{dA}
\end{equation}
The next step is to develope an iterative scheme which will give the first
terms of the expansion, in powers of covariant or ordinary derivatives of
the relevant operator.

\section{Generalized method}

In the previous section we described the first common steps between the
derivative expansion method developed previously \cite{moss}, \cite{moss1},
for operators of the form $\Delta =-\nabla ^{2}+X$ , and the generalized
method presented here. The old method is not applicable in the case of some
quadratic operators we encounter in gauge field theories. The reason is that
in different gauges than the Feynman one, we get operators of the following
form 
\begin{equation}
\Delta =-\delta _{\mu }^{\nu }\nabla ^{2}+(1-\xi ^{-1})\nabla _{\mu }\nabla
^{\nu }+X\delta _{\mu }^{\nu }
\end{equation}
The material below is a generalization of the derivative expansion method in
order to handle the quadratic operators in 't Hooft gauge fixing. We avoid
describing in great detail the derivation of the various expressions. The
reason is that the algebra is too lengthy to be presented.

The heat kernel equation of the above operator can be written in the
momentum space, after a Taylor expansion of $X(k,{\bf x})$ , in powers of $%
{\bf x-x}^{\prime }\equiv x^{\mu }-x^{\prime \mu }$ around $X(k,{\bf x}%
^{\prime }),$ as follows 
\begin{equation}
\left[ \delta _{\mu }^{\nu }k^{2}-(1-\xi ^{-1})k_{\mu }k^{\nu }+\delta _{\mu
}^{\nu }\sum_{r=0}^{\infty }\frac{1}{r\,!}X_{,\mu _{1}...\mu _{r}}\delta
^{\mu _{1}}...\delta ^{\mu r}\right] K(k,{\bf x}^{\prime },t)=-\stackrel{.}{K%
}(k,{\bf x}^{\prime },t)  \label{kk}
\end{equation}
where $\delta ^{\mu }=i\,\partial /\partial k^{\mu }$ is the transformed $%
x^{\mu }-x^{\prime \mu }$ . We define 
\begin{equation}
P_{\mu }^{\nu }=\delta _{\mu }^{\nu }-\widehat{k}_{\mu }\widehat{k}^{\nu
}\qquad Q_{\mu }^{\nu }=\widehat{k}_{\mu }\widehat{k}^{\nu }
\end{equation}
where 
\begin{equation}
\widehat{k}_{\mu }\widehat{k}^{\nu }=\frac{k_{\mu }k^{\nu }}{k^{2}}
\end{equation}
It is obvious that 
\begin{equation}
PQ\equiv P_{\mu }^{\lambda }Q_{\lambda }^{\nu }=0\qquad P_{\mu }^{\nu
}+Q_{\mu }^{\nu }=\delta _{\mu }^{\nu }
\end{equation}
Performing the derivative expansion approximation we find from Eq. \ref{kk},
at zero order 
\begin{equation}
K_{0}=e^{-\left[ \left( P_{\mu }^{\nu }+\xi ^{-1}Q_{\mu }^{\nu }\right)
k^{2}+\delta _{\mu }^{\nu }X_{0}\right] t}=\left( Pe^{-k^{2}t}+Qe^{-\xi
^{-1}k^{2}t}\right) e^{-X_{0}t}  \label{ko}
\end{equation}
$K_{0}$ can further be written in terms of the eigenvalues $m_{i}^{2}$ of
the matrix $X_{0}$ and the matrices $T_{i},$ which are defined by 
\begin{equation}
X_{0}=\sum m_{i}^{2}T_{i}  \label{X0}
\end{equation}
Using expression \ref{ko}, a recursion equation can be obtained 
\begin{equation}
\stackrel{.}{a}_{n}=-\sum\limits_{0<r+s\leq n}\sum\limits_{i,j}\frac{1}{r\,!}%
T_{i}X_{s,\mu _{1}...\mu _{r}}T_{j}e^{-\left( m_{j}^{2}-m_{i}^{2}\right) t}\,%
\widehat{D}^{\mu _{1}}...\widehat{D}^{\mu _{r}}a_{n-r-s}\;.  \label{iter}
\end{equation}
where 
\begin{equation}
\widehat{D}^{\rho }=\delta _{\mu }^{\nu }i\frac{\partial }{\partial k^{\rho }%
}-\Gamma _{\mu }^{\nu \rho }=\delta _{\mu }^{\nu }\;\delta ^{\rho }-\Gamma
_{\mu }^{\nu \rho }
\end{equation}
The connection term is 
\begin{equation}
\Gamma _{\mu }^{\nu \rho }\equiv \Gamma ^{\rho }=2ik^{\rho }t\left( P+\xi
^{-1}Q\right) -iPP^{,\rho }\left( 1-e^{\left( 1-\xi ^{-1}\right)
k^{2}t}\right) -iQP^{,\rho }\left( -1+e^{\left( -1+\xi ^{-1}\right)
k^{2}t}\right)
\end{equation}
where 
\begin{equation}
P^{,\rho }\equiv P_{\mu }^{\nu ,\rho }=\frac{\partial P_{\mu }^{\nu }}{%
\partial k^{\rho }}
\end{equation}
The following relations are very helpful to proceed further into the algebra
of the iterative scheme.

{\it Zeroth order :} 
\begin{equation}
\text{\qquad }PP=P.\qquad QQ=Q\quad .
\end{equation}

{\it First order :} 
\begin{equation}
PP^{,\rho }+P^{,\rho }P=P^{,\rho }.\qquad {\rm tr}\left[ PP^{,\rho }\right] =%
{\rm tr}\left[ P^{,\rho }\right] =0\qquad
\end{equation}

{\it Second order :} 
\begin{equation}
{\rm tr}\left[ P^{,\rho }P^{,\sigma }\right] =\frac{2}{k^{2}}P^{\rho \sigma
}.\qquad {\rm tr}\left[ PP^{,\rho }P^{,\sigma }\right] =\frac{1}{k^{2}}%
P^{\rho \sigma }
\end{equation}
also 
\begin{equation}
{\rm tr}\left[ P^{,\rho \sigma }\right] =0.\qquad {\rm tr}\left[ PP^{,\rho
\sigma }\right] =-\frac{2}{k^{2}}P^{\rho \sigma }.\qquad {\rm tr}\left[
PP^{,\rho }PP^{,\sigma }\right] =0
\end{equation}
Terms including $Q$ are 
\begin{equation}
P^{,\rho }=-Q^{,\rho }.\qquad P^{,\rho \sigma }=-Q^{,\rho \sigma }
\end{equation}
and 
\begin{equation}
{\rm tr}\left[ QP^{,\rho }\right] =0.\qquad {\rm tr}\left[ QP^{,\rho
}P^{,\sigma }\right] =\frac{1}{k^{2}}P^{\rho \sigma }.\qquad {\rm tr}\left[
QP^{,\rho \sigma }\right] =\frac{2}{k^{2}}P^{\rho \sigma }
\end{equation}
\begin{equation}
{\rm tr}\left[ PP^{,\rho }QP^{,\sigma }\right] =\frac{1}{k^{2}}P^{\rho
\sigma }.\qquad {\rm tr}\left[ QP^{,\rho }QP^{,\sigma }\right] =0
\end{equation}
For the terms containing $\Gamma $ we find that 
\begin{equation}
{\rm tr}\left[ P\,\Gamma ^{,\rho }\Gamma ^{,\sigma }\right] =-4k^{\rho
}k^{\sigma }t^{2}({\rm tr}P)-\frac{2}{k^{2}}P^{\rho \sigma }\left[ \cosh
\left( k^{2}(-1+\xi ^{-1})t\right) -1\right]
\end{equation}
\begin{equation}
{\rm tr}\left[ P\,\Gamma ^{,\rho \sigma }\right] =2i\delta ^{\rho \sigma
}t\,({\rm tr}P)-i\frac{1}{k^{2}}P^{\rho \sigma }\left[ e^{k^{2}(1-\xi
^{-1})t}-1\right] +i\frac{1}{k^{2}}P^{\rho \sigma }\left[ e^{k^{2}(-1+\xi
^{-1})t}-1\right]
\end{equation}
for the matrix $Q$%
\begin{equation}
{\rm tr}\left[ Q\,\Gamma ^{,\rho }\Gamma ^{,\sigma }\right] =-4k^{\rho
}k^{\sigma }t^{2}\xi ^{-2}-\frac{2}{k^{2}}P^{\rho \sigma }\left[ \cosh
\left( k^{2}(-1+\xi ^{-1})t\right) -1\right]
\end{equation}
\begin{equation}
{\rm tr}\left[ Q\,\Gamma ^{,\rho \sigma }\right] =2i\delta ^{\rho \sigma
}t\,\xi ^{-1}+i\frac{1}{k^{2}}P^{\rho \sigma }\left[ e^{k^{2}(1-\xi
^{-1})t}-1\right] -i\frac{1}{k^{2}}P^{,\rho \sigma }\left[ e^{k^{2}(\xi
^{-1}-1)t}-1\right]
\end{equation}

The recursion equation \ref{iter}, leads to an iterative scheme with which
we can evaluate the first terms in powers of ordinary derivatives of the
heat kernel. In the first iteration we get 
\begin{equation}
\stackrel{.}{a_{1}}=\Gamma ^{\rho }\sum_{i,j}T_{i}X_{0,\rho
}T_{j}\,e^{\left( m_{i}^{2}-m_{j}^{2}\right)
t}-\sum_{i,j}T_{i}X_{1}T_{j}\,e^{\left( m_{i}^{2}-m_{j}^{2}\right) t}
\end{equation}
and after an integration from $t=0$ to $t=\infty $ we find 
\begin{eqnarray}
a_{1} &=&2ik^{\rho }\left( P+\xi ^{-1}Q\right) \sum_{i,j}T_{i}X_{0,\rho
}T_{j}\,f_{ij}-\sum_{i,j}T_{i}X_{1}T_{j}\,g_{ij}-  \nonumber \\
&&\ -i\left( P-Q\right) P^{,\rho }\sum_{i,j}T_{i}X_{0,\rho }T_{j}\,g_{ij}+ 
\nonumber \\
&&\ +i\sum_{i,j}T_{i}X_{0,\rho }T_{j}\,\left[ 
\begin{array}{c}
PP^{,\rho }g\left( m^{2}+\left( 1-\xi ^{-1}\right) k^{2}\right) \\ 
-QP^{,\rho }g\left( m^{2}-\left( 1-\xi ^{-1}\right) k^{2}\right)
\end{array}
\right]
\end{eqnarray}
where 
\begin{equation}
f(m^{2})\equiv f_{ij}(t)=\frac{\partial }{\partial m^{2}}g_{ij}(t),\qquad
m^{2}=m_{i}^{2}-m_{j}^{2}
\end{equation}
\begin{equation}
g(m^{2})\equiv g_{ij}(t)=m^{-2}\left( e^{m^{2}t}-1\right)
\end{equation}
The lengthy evaluation of the second term $\stackrel{.}{a}_{2}$ is described
in Appendix B.

Finally after considerable cancellations, we find through the expression Eq. 
\ref{dA} that for a 't Hooft gauge fixing the first terms of the derivative
expansion are: 
\begin{equation}
A_{0}=(4\pi t)^{d/2}\left( {\rm tr}P+\xi ^{d/2}\right) K(t)\,\sum_{i}{\rm tr}%
\left( T_{i}\right) \,e^{-m_{i}^{2}t}  \label{onet}
\end{equation}
where 
\begin{equation}
K(t)=\int d\mu (k)\,\,e^{-k^{2}t}
\end{equation}
The term $A_{1}$ does not contribute in the effective action, in our case,
because leads to a total divergence. 
\begin{eqnarray}
A_{2}/(4\pi t)^{d/2} &=&-\frac{1}{6}t^{2}\left( {\rm tr}P+\xi
^{-1+d/2}\right) \sum_{i}{\rm tr}\left( T_{i}\nabla _{\mu }\nabla ^{\mu
}X_{0}T_{i}\right) \,\,K_{i}(t)-  \nonumber \\
&&\ \ -\frac{2}{3}t\sum_{i}{\rm tr}\left( T_{i}\nabla _{\mu }\nabla ^{\mu
}X_{0}T_{i}\right) \,\,\left[ \widetilde{K}_{i}(t)-K_{i}^{(\xi )}(t)\right] +
\nonumber \\
&&\ \ +\left( {\rm tr}P+\xi ^{-1+d/2}\right) \,K(t)\,\sum_{i,j}{\rm tr}%
\left( T_{i}\nabla _{\mu }X_{0}T_{j}\nabla ^{\mu }X_{0}T_{i}\right) \eta
_{ij}(t)+  \nonumber \\
&&\ \ +\frac{2}{3}\left( \widetilde{K}(t)-K^{(\xi )}(t)\right) \sum_{i,j}%
{\rm tr}\left( T_{i}\nabla _{\mu }X_{0}T_{j}\nabla ^{\mu }X_{0}T_{i}\right)
\chi _{ij}(t)+  \nonumber \\
&&\ \ +K(t)\,\left( {\rm tr}P+\xi ^{d/2}\right) \sum_{i,j}{\rm tr}\left(
T_{i}X_{1}T_{j}X_{1}T_{i}\right) \chi _{ij}(t)+  \nonumber \\
&&\ \ +\frac{2}{3}\sum_{i,j}{\rm tr}\left( T_{i}\nabla _{\mu
}X_{0}T_{j}\nabla ^{\mu }X_{0}T_{i}\right) \left[ 
\begin{array}{c}
t\,\lambda _{ij}(t)+\mu _{ij}(t)+ \\ 
+\frac{\xi }{\xi -1}\nu _{ij}(t)
\end{array}
\right]  \label{twot}
\end{eqnarray}
where 
\begin{equation}
K_{i}(t)=\int_{-\infty }^{\infty }d\mu
(k)\,\,e^{-k^{2}t}e^{-m_{i}^{2}t}=K(t)\,e^{-m_{i}^{2}t}
\end{equation}
\begin{equation}
\widetilde{K}_{i}(t)=\int_{-\infty }^{\infty }d\mu (k)\,\,\frac{1}{k^{2}}%
e^{-k^{2}t}e^{-m_{i}^{2}t}=\widetilde{K}(t)\,e^{-m_{i}^{2}t}
\end{equation}
and 
\begin{eqnarray}
\lambda _{ij}(t) &=&\frac{1}{2}\int d\mu (k)\frac{k^{-2}}{-m^{2}+k^{2}(1-\xi
^{-1})}\theta _{ij}(t)+  \nonumber \\
&&\ \ +\frac{1}{2}\int d\mu (k)\frac{k^{-2}}{m^{2}+k^{2}(1-\xi ^{-1})}\theta
_{ji}(t)
\end{eqnarray}
\begin{eqnarray}
\mu _{ij}(t) &=&\frac{1}{2}\frac{1}{m^{2}}\int d\mu (k)\frac{k^{-2}}{%
-m^{2}+k^{2}(1-\xi ^{-1})}\left[ \zeta _{ij}(t)-\zeta _{ji}(t)\right] + 
\nonumber \\
&&\ \ +\frac{1}{2}\frac{1}{m^{2}}\int d\mu (k)\frac{k^{-2}}{%
m^{2}+k^{2}(1-\xi ^{-1})}\left[ \zeta _{ij}(t)-\zeta _{ji}(t)\right]
\end{eqnarray}
\begin{eqnarray}
\nu _{ij}(t) &=&\frac{1}{2}\int d\mu (k)\frac{k^{-4}}{-m^{2}+k^{2}(1-\xi
^{-1})}\left[ -\zeta _{ij}(t)+\zeta _{ji}(t)\right] +  \nonumber \\
&&\ \ \ \ +\frac{1}{2}\int d\mu (k)\frac{k^{-4}}{m^{2}+k^{2}(1-\xi ^{-1})}%
\left[ \zeta _{ij}(t)-\zeta _{ji}(t)\right]
\end{eqnarray}
also 
\begin{equation}
\eta _{ij}(t)=-m^{-6}\left( 1+\frac{1}{6}m^{4}t^{2}\right) \left(
e^{-m_{i}^{2}t}-e^{-m_{j}^{2}t}\right) -\frac{1}{2}m^{-4}t\left(
e^{-m_{i}^{2}t}+e^{-m_{j}^{2}t}\right)
\end{equation}
\begin{equation}
\chi _{ij}(t)=-\frac{1}{2}m^{-2}t\left(
e^{-m_{i}^{2}t}-e^{-m_{j}^{2}t}\right)
\end{equation}
In the case that $i=j$%
\begin{equation}
\eta _{ii}(t)=\frac{1}{12}t^{3}e^{-m_{i}^{2}t}\ ,\qquad \chi _{ii}(t)=\frac{1%
}{2}t^{2}e^{-m_{i}^{2}t}
\end{equation}
The above results in the Feynman gauge $\xi =1$ , reduce to the respective
expressions, in \cite{moss1}. We should ignore the spacetime indices if the
operators do not carry such indices, otherwise we should take in account
them, assuming that the traces in the respective equations in \cite{moss1},
are also over spacetime indices.

The above developed expansion formulae can also be given in a different and
more lengthy in our case form, using the covariant derivatives ${\cal D}%
_{\mu }$ instead of the simple derivatives $\nabla _{\mu }$ . The above
expansion can be carefully transformed in the covariant form in a similar
way with that proposed in \cite{moss1}.

Note that 
\[
\text{For }\xi =1\qquad \widetilde{K}(t)=K^{(\xi )}(t)\qquad \text{and\qquad 
}\lambda _{ij}(t)=\mu _{ij}(t)=\nu _{ij}(t)=0 
\]
and 
\[
\text{For }\xi =0\qquad K^{(\xi )}(t)=0\qquad \text{and\qquad }\lambda
_{ij}(t)=\mu _{ij}(t)=\nu _{ij}(t)=0 
\]
Note that one can prove, in order to test the validity of the developed
expansion, that the derivative of the corrected effective action with
respect to the $\xi $ parameter, evaluated at $\xi =1$, is zero. Thus small
deviations from the Feynman gauge leave the calculated corrected action
invariant, as one should expect.

\subsection{Landau gauge}

The derivative expansion in the Landau gauge $\xi =0$ reduces to the
following equations : 
\begin{equation}
A_{0}=(4\pi t)^{d/2}\left( {\rm tr}P\right) K(t)\,\sum_{i}{\rm tr}\left(
T_{i}\right) \,e^{-m_{i}^{2}t}
\end{equation}
\begin{eqnarray}
A_{2}/(4\pi t)^{d/2} &=&-\frac{1}{6}t^{2}\left( {\rm tr}P\right) \sum_{i}%
{\rm tr}\left( T_{i}\nabla _{\mu }\nabla ^{\mu }X_{0}T_{i}\right)
\,\,K_{i}(t)-  \nonumber \\
&&\ \ \ -\frac{2}{3}t\sum_{i}{\rm tr}\left( T_{i}\nabla _{\mu }\nabla ^{\mu
}X_{0}T_{i}\right) \,\,\widetilde{K}_{i}(t)+  \nonumber \\
&&\ \ \ +\left( {\rm tr}P\right) \,K(t)\,\sum_{i,j}{\rm tr}\left(
T_{i}\nabla _{\mu }X_{0}T_{j}\nabla ^{\mu }X_{0}T_{i}\right) \eta _{ij}(t)+ 
\nonumber \\
&&\ \ \ +\frac{2}{3}\widetilde{K}(t)\sum_{i,j}{\rm tr}\left( T_{i}\nabla
_{\mu }X_{0}T_{j}\nabla ^{\mu }X_{0}T_{i}\right) \chi _{ij}(t)+  \nonumber \\
&&\ \ \ +K(t)\,\left( {\rm tr}P\right) \sum_{i,j}{\rm tr}\left(
T_{i}X_{1}T_{j}X_{1}T_{i}\right) \chi _{ij}(t)
\end{eqnarray}

\subsection{Finite temperature corrections}

In some stages of the early Universe evolution, we often expect the
particles to be in thermal equilibrium. Therefore a field $\Phi $ can be in
contact with this heat bath. The result is that it gets a temperature
dependent mass. Since we deal with an ensemble of particles in thermal
equilibrium, the background state in which we perform calculations is no
longer the ground state of the Hamiltonian but a thermal bath at temperature 
$T$. To study this situation is essential to use a statistical quantity, the
finite temperature effective potential $V(\Phi ,T)$. This is the free energy
density associated with the $\Phi $ field. At $T\neq 0$ quantum statistics
is equivalent to Euclidean quantum field theory in a space which is periodic
with period $\beta =1/T$ along the ``imaginary time`` axis. Thus we can
write the partition function $Z$ as 
\begin{eqnarray}
Z &=&\sum_{\Phi }<\Phi ({\bf x}),t=0\,\mid {\bf e}^{-\beta \,H}\mid \,\Phi (%
{\bf x}),t=0{\bf >}  \nonumber \\
&\propto &\int {\cal D}[\Phi \,]\exp \{\int_{0}^{\beta }d\tau \int d^{3}x%
{\cal \,L}\}
\end{eqnarray}
where the integral is periodic with period depending on the kind of
statistics the particles obey 
\begin{equation}
bosons:\,\;\phi (\tau =0,{\bf x})=\phi (\tau =\beta ,{\bf x})
\end{equation}
\begin{equation}
fermions:\,\;\psi (\tau =0,{\bf x})=-\psi (\tau =\beta ,{\bf x})
\end{equation}

The fields here represent ensemble of particles in local thermal
equilibrium. These thermal averages vary over space. The short discussion we
had, implies that the integral measure, in momentum space, becomes 
\begin{equation}
\int d\mu (k)\,f(k^{i},k^{0})\rightarrow \sum_{n=-\infty }^{\infty }\beta
^{-1}\int \frac{d^{3}k}{\left( 2\pi \right) ^{3}}\,f\left( k^{i},\frac{2\pi
n+s\pi }{\beta }\right)
\end{equation}
where $s=0$ for bosons and $s=1$ for fermions. We need to calculate now the $%
\zeta ^{\prime }({\bf x,}0)$ function in order to find the one loop
corrections through the expression Eq. \ref{det}. The $\zeta ^{\prime }({\bf %
x,}s)$ can be written as the sum of terms $\zeta _{n}^{\prime }({\bf x,}s)$
with different number $n$ of derivatives. 
\begin{equation}
\zeta _{n}^{\prime }({\bf x,}s)=\frac{1}{\Gamma (s)}\int_{0}^{\infty
}dt\,t^{s-1}{\rm tr}[K_{0}a_{n}]
\end{equation}
The above expression together with the developed expansion, see Eq. \ref
{onet}, \ref{twot}, contains all the information needed to find quantum
corrections at finite temperature. The one loop contribution, at finite
temperature, is given by the following expression 
\begin{equation}
\Gamma ^{(1)}={\frac{1}{T}}\int d^{3}x\,\,\left[ -{\frac{1}{2}}\zeta
^{^{\prime }}(x,0)\right]
\end{equation}
It is straightforward to find, but lengthy to present, the temperature
corrected expansion in 't Hooft gauge. In the previous work \cite{moss1} the
temperature corrected expansion was given in Feynman gauge while here is
given, for comparison, in Landau gauge. The way we can calculate this
expansion for bosons and fermions is described in appendix A. 
\begin{eqnarray}
{\zeta }^{\prime }{_{2}(x,0)} &=&\left( {\rm tr}P\right)
\sum_{i}tr[T_{i}\nabla _{\mu }\nabla ^{\mu }X_{0}T_{i}]\rho _{i}^{\prime
}(0)+  \nonumber \\
&&+\sum_{i}tr[T_{i}\nabla _{\mu }\nabla ^{\mu }X_{0}T_{i}]\widetilde{\rho }%
_{i}^{\prime }(0)+  \nonumber \\
&&+\left( {\rm tr}P\right) \sum_{ij}tr[T_{i}\nabla _{\mu }X_{0}T_{j}\nabla
^{\mu }X_{0}T_{i}]\eta _{ij}^{\prime }(0)+  \nonumber \\
&&+\sum_{ij}tr[T_{i}\nabla _{\mu }X_{0}T_{j}\nabla ^{\mu }X_{0}T_{i}]%
\widetilde{\chi }_{ij}^{\prime }(0)+  \nonumber \\
&&+\left( {\rm tr}P\right) \sum_{ij}tr[T_{i}X_{1}T_{j}X_{1}T_{i}]\chi
_{ij}^{\prime }(0).  \label{kinT}
\end{eqnarray}
The new functions, we introduce, are given below, in the high temperature ($%
m/T<1$) regime and for bosons, 
\begin{equation}
\rho _{i}^{\prime }(0)\simeq -{\frac{T}{48\pi }}m_{i}^{-1}
\end{equation}
\begin{equation}
\widetilde{\rho }_{i}^{\prime }(0)\simeq -{\frac{T}{6\pi }}m_{i}^{-1}
\end{equation}
\begin{equation}
\eta _{ij}^{\prime }(0)\simeq {\frac{T}{192\pi }}\,{\frac{4\left(
m_{i}^{2}+m_{j}^{2}\right) }{m_{i}m_{j}(m_{i}+m_{j})^{3}}}
\end{equation}
\begin{equation}
\chi _{ij}^{\prime }(0)\simeq {\frac{T}{8\pi }}(m_{i}+m_{j})^{-1}
\end{equation}
\begin{equation}
\widetilde{\chi }_{ij}^{\prime }(0)\simeq {\frac{T}{12\pi }}\frac{1}{%
m_{i}m_{j}}(m_{i}+m_{j})^{-1}
\end{equation}

One should note here that, the finite temperature effective action is not
well defined because of the nonanalytic behavior of the two point functions
involved. This problem becomes manifest from the different results we get,
taking different order of zero limits of the four external momenta \cite{das}%
. However it can give a correct approximate estimation of the critical
temperature. As far as the nature of the transition is concerned, something
that would be crucial for the cosmological electroweak phase transition \cite
{zarikas}, the various proposed improved methods do not seem to agree \cite
{asprouli}.

\section{Scalar Electrodynamics}

In this section we will apply as an example, the derivative expansion method
we just described, in the toy model of scalar electrodynamics. For loop
corrections of this model see \cite{fra}, \cite{ja}.

The Lagrangian can be written as the sum of the physical and the ghost
contributions. 
\begin{equation}
{\cal L}={\cal L}_{phys}+{\cal L}_{ghost}
\end{equation}
This model just has a charged scalar field and an electromagnetic one. 
\begin{eqnarray}
{\cal L}_{phys} &=&-(D_\mu \Phi )(D^\mu \Phi )^{*}-\mu ^2\Phi \Phi ^{*}-%
\frac 14\lambda (\Phi \Phi ^{*})^2-\frac 14F_{\mu \nu }F^{\mu \nu } 
\nonumber \\
&&-\frac 1{2\xi }\left( \partial _\mu A^\mu -\xi q\phi _{cl}\,\phi _1\right)
^2  \label{phys}
\end{eqnarray}
where 
\begin{equation}
\Phi =\frac 1{\sqrt{2}}\left( \,\phi _{cl}+\,\phi _1+i\,\phi _2\right)
\end{equation}
The U(1) covariant derivative is 
\begin{equation}
D_\mu =\partial _\mu +iqA_\mu
\end{equation}
and the electromagnetic tensor 
\begin{equation}
F_{\mu \nu }=\partial _\mu A_\nu -\partial _\nu A_\mu
\end{equation}

Since we want to find the quadratic part of the Lagrangian we expand the
expression Eq. \ref{phys} and drop total divergences$.$ We get 
\begin{eqnarray}
{\cal L}_{quadr} &=&-\frac{1}{2}\partial _{\mu }\phi _{1}\partial ^{\mu
}\phi _{1}-\frac{1}{2}\left( \mu ^{2}+\frac{3}{4}\lambda \phi
_{cl}^{2}\right) \phi _{1}^{2}+  \nonumber \\
&&\ \ \ \ \ \ \ \ \ -\frac{1}{2}\partial _{\mu }\phi _{2}\partial ^{\mu
}\phi _{2}-\frac{1}{2}\left( \mu ^{2}+\frac{1}{4}\lambda \phi _{cl}^{2}+\xi
q^{2}\phi _{cl}^{2}\,\right) \phi _{2}^{2}-  \nonumber \\
&&\ \ \ \ \ \ \ \ \ -\frac{1}{2}\left[ (1-\xi ^{-1})A^{\mu }\partial _{\mu
}\partial ^{\nu }A_{\nu }-A^{\mu }\delta _{\mu }^{\nu }\partial _{\lambda
}\partial ^{\lambda }A_{\nu }+q^{2}\phi _{cl}^{2}A^{\mu }A_{\mu }\right] - 
\nonumber \\
&&\ \ \ \ \ \ \ \ \ -2iq\phi _{2}A^{\mu }\partial _{\mu }\phi _{cl}+%
\overline{c}\left[ -\partial ^{\mu }\partial _{\mu }+\xi q^{2}\phi
_{cl}^{2}\right] c
\end{eqnarray}
It can be written as 
\begin{equation}
{\cal L}_{quadr}=-\frac{1}{2}\eta _{\nu }^{\intercal }\,\Delta _{\mu }^{\nu
}\eta ^{\mu }+\overline{c\,}\,\Delta _{ghost}c
\end{equation}
where 
\begin{equation}
\eta _{\nu }^{\intercal }=\left( 
\begin{array}{ccc}
A_{\nu } & \phi _{1} & \phi _{2}
\end{array}
\right)
\end{equation}
and 
\begin{equation}
\Delta _{\mu }^{\nu }=\left( 
\begin{array}{ccc}
\Delta _{\mu \left( A\right) }^{\nu } & 0 & X_{1\mu } \\ 
0 & \Delta _{11}^{\phi } & 0 \\ 
X_{1}^{\nu } & 0 & \Delta _{22}^{\phi }
\end{array}
\right)
\end{equation}
The off diagonal entries are the non zero kinetic terms. 
\begin{equation}
X_{1\mu }=2iq\,\partial _{\mu }\phi _{cl}\,
\end{equation}
The fluctuation operators are 
\begin{equation}
\Delta _{\phi }=-\nabla ^{2}+X_{\phi }
\end{equation}
with 
\begin{equation}
X_{\phi }=\left( 
\begin{array}{cc}
\mu ^{2}+\frac{3}{4}\lambda \phi _{cl}^{2} & 0 \\ 
0 & \mu ^{2}+\frac{1}{4}\lambda \phi _{cl}^{2}+\xi q^{2}\phi _{cl}^{2}
\end{array}
\right)
\end{equation}
The gauge field operator is 
\begin{equation}
\Delta _{\mu \left( A\right) }^{\nu }=-\delta _{\mu }^{\nu }\nabla
^{2}+(1-\xi ^{-1})\nabla _{\mu }\nabla ^{\nu }+q^{2}\phi _{cl}^{2}\delta
_{\mu }^{\nu }
\end{equation}
and the ghost one 
\begin{equation}
\Delta _{gh}=-\nabla ^{2}+\xi q^{2}\phi _{cl}^{2}
\end{equation}
So the quantities we need for the formalism of the derivative expansion, are 
\begin{equation}
X_{0}=\left( 
\begin{array}{ccc}
\delta _{\mu }^{\nu }m_{1}^{2} & 0 & 0 \\ 
0 & m_{2}^{2} & 0 \\ 
0 & 0 & m_{3}^{2}
\end{array}
\right)
\end{equation}
\begin{equation}
X_{1}=\left( 
\begin{array}{ccc}
0 & 0 & X_{1\mu } \\ 
0 & 0 & 0 \\ 
X_{1}^{\nu } & 0 & 0
\end{array}
\right)
\end{equation}
with masses 
\begin{equation}
m_{1}^{2}=q^{2}\phi _{cl}^{2}\qquad m_{2}^{2}=\mu ^{2}+\frac{3}{4}\lambda
\phi _{cl}^{2}\qquad m_{3}^{2}=\mu ^{2}+\frac{1}{4}\lambda \phi
_{cl}^{2}+\xi q^{2}\phi _{cl}^{2}\qquad
\end{equation}
The projection matrices are 
\begin{equation}
T_{1}=\left( 
\begin{array}{ccc}
1 & 0 & 0 \\ 
0 & 0 & 0 \\ 
0 & 0 & 0
\end{array}
\right) \;,\qquad T_{2}=\left( 
\begin{array}{ccc}
0 & 0 & 0 \\ 
0 & 1 & 0 \\ 
0 & 0 & 0
\end{array}
\right) \;,\qquad T_{3}=\left( 
\begin{array}{ccc}
0 & 0 & 0 \\ 
0 & 0 & 0 \\ 
0 & 0 & 1
\end{array}
\right) \;
\end{equation}
Since now the matrix $X$ is 
\begin{equation}
X=\left( 
\begin{array}{ccc}
\delta _{\mu }^{\nu }m_{1}^{2} & 0 & X_{1\mu } \\ 
0 & m_{2}^{2} & 0 \\ 
X_{1}^{\nu } & 0 & m_{3}^{2}
\end{array}
\right)
\end{equation}
it is trivial to see that here we should introduce the following definitions 
\begin{equation}
\Pi _{\mu }^{\nu }=\left( 
\begin{array}{ccc}
P_{\mu }^{\nu } & 0 & 0 \\ 
0 & 1 & 0 \\ 
0 & 0 & 1
\end{array}
\right) =P_{\mu }^{\nu }T_{1}+T_{2}+T_{3}
\end{equation}
\begin{equation}
\Theta _{\mu }^{\nu }=\left( 
\begin{array}{ccc}
Q_{\mu }^{\nu } & 0 & 0 \\ 
0 & 0 & 0 \\ 
0 & 0 & 0
\end{array}
\right) =Q_{\mu }^{\nu }T_{1}
\end{equation}
It is obvious that $\Pi _{\mu }^{\nu }$ and $\Theta _{\mu }^{\nu }$ satisfy
similar expressions to $P_{\mu }^{\nu }$ and $Q_{\mu }^{\nu }$ . 
\begin{equation}
\Pi _{\mu }^{\lambda }\Theta _{\lambda }^{\nu }=0\;,\qquad \Pi _{\mu }^{\nu
}+\Theta _{\mu }^{\nu }=\delta _{\mu }^{\nu }T_{1}+T_{2}+T_{3}
\end{equation}
\begin{equation}
\Pi _{\mu }^{\lambda }\Pi _{\lambda }^{\nu }=\Pi {}_{\mu }^{\nu }\qquad
\Theta _{\mu }^{\lambda }\Theta _{\lambda }^{\nu }=\Theta {}_{\mu }^{\nu
}\quad .
\end{equation}
The important expressions are, 
\begin{eqnarray}
{\rm tr}\left[ \Pi \,\Gamma ^{,\rho \sigma }\right] &=&2i\delta ^{\rho
\sigma }t\,\left[ {\rm tr}(P\,)T_{1}+T_{2}+T_{3}\right] -i\frac{1}{k^{2}}%
P^{,\rho \sigma }T_{1}\left[ e^{k^{2}(1-\xi ^{-1})t}-1\right] +  \nonumber \\
&&\ \ \ \ \ \ +i\frac{1}{k^{2}}P^{,\rho \sigma }T_{1}\left[ e^{k^{2}(-1+\xi
^{-1})t}-1\right]
\end{eqnarray}
while for the matrix $\Theta {}_{\mu }^{\nu }$ , 
\begin{equation}
{\rm tr}\left[ \Theta \,\Gamma ^{,\rho }\Gamma ^{,\sigma }\right] =-4k^{\rho
}k^{\sigma }t^{2}\xi ^{-2}\,T_{1}-\frac{2}{k^{2}}P^{,\rho \sigma
}T_{1}\left[ \cosh \left( k^{2}(-1+\xi ^{-1})t\right) -1\right] \;.
\end{equation}
Proceeding as before, the first terms of the derivative expansion are: 
\begin{equation}
A_{0}=(4\pi t)^{d/2}K(t)\,\sum_{i}{\rm tr}\left( T_{i}\right)
\,e^{-m_{i}^{2}t}\left( \delta _{i1}{\rm tr}P\,+\delta _{i2}+\delta
_{i3}+\delta _{i1}\xi ^{d/2}\right)
\end{equation}
\begin{eqnarray}
A_{2}/(4\pi t)^{d/2} &=&-\frac{1}{6}t^{2}\sum_{i}{\rm tr}\left( T_{i}\nabla
_{\mu }\nabla ^{\mu }X_{0}T_{i}\right) \left( \delta _{i1}{\rm tr}P\,+\delta
_{i2}+\delta _{i3}+\delta _{i1}\xi ^{-1+d/2}\right) \,\,K_{i}(t)-  \nonumber
\\
&&\ \ \ \ \ -\frac{2}{3}t\sum_{i}{\rm tr}\left( T_{i}\nabla _{\mu }\nabla
^{\mu }X_{0}T_{i}\right) \delta _{i1}\,\left[ \,\widetilde{K}%
_{i}(t)-K_{i}^{(\xi )}(t)\right]  \nonumber \\
&&\ \ \ \ \ +\,K(t)\,\sum_{i}{\rm tr}\left( T_{i}\nabla _{\mu
}X_{0}T_{i}\nabla ^{\mu }X_{0}T_{i}\right) \left[ 
\begin{array}{c}
\delta _{i1}{\rm tr}P\,+\delta _{i2}+\delta _{i3} \\ 
+\delta _{i1}\xi ^{-1+d/2}
\end{array}
\right] \eta _{ii}(t)+  \nonumber \\
&&\ \ \ \ \ +\frac{2}{3}\left( \widetilde{K}(t)-K^{(\xi )}(t)\right)
\sum_{i,j}{\rm tr}\left( T_{i}\nabla _{\mu }X_{0}T_{j}\nabla ^{\mu
}X_{0}T_{i}\right) \delta _{i1}\chi _{ij}(t)+  \nonumber \\
&&\ \ \ \ \ +K(t)\,\sum_{i,j}{\rm tr}\left( T_{i}X_{1}T_{j}X_{1}T_{i}\right)
\left( \delta _{i1}+\delta _{j1}\right) \left[ {\rm tr}P\,+\xi ^{d/2}\right]
\chi _{ij}(t)+  \nonumber \\
&&\ \ \ \ \ +\frac{2}{3}\sum_{i,j}{\rm tr}\left( T_{i}\nabla _{\mu
}X_{0}T_{j}\nabla ^{\mu }X_{0}T_{i}\right) \left[ 
\begin{array}{c}
t\,\lambda _{ij}(t)+\mu _{ij}(t)+ \\ 
+\frac{\xi }{\xi -1}\nu _{ij}(t)
\end{array}
\right] \delta _{i1}
\end{eqnarray}

The only non zero combinations are 
\begin{eqnarray}
T_{1}\partial _{\mu }X_{0}T_{1}\partial ^{\mu }X_{0}T_{1} &=&\left( \partial
^{\mu }m_{1}^{2}\right) ^{2}T_{1},\qquad T_{2}\partial _{\mu
}X_{0}T_{2}\partial ^{\mu }X_{0}T_{2}=\left( \partial ^{\mu
}m_{2}^{2}\right) ^{2}T_{2}\quad  \nonumber \\
T_{3}\partial _{\mu }X_{0}T_{3}\partial ^{\mu }X_{0}T_{3} &=&\left( \partial
^{\mu }m_{3}^{2}\right) ^{2}T_{3}\qquad \quad \quad \quad
\end{eqnarray}
and 
\begin{equation}
T_{1}X_{1}T_{3}X_{1}T_{1}=-4q^{2}\,\partial _{\mu }\phi _{cl}\,\partial
^{\nu }\phi _{cl}\,T_{1},\qquad T_{3}X_{1}T_{1}X_{1}T_{3}=-4q^{2}\,\partial
_{\mu }\phi _{cl}\,\partial ^{\nu }\phi _{cl}\,T_{3}\quad \quad \quad
\end{equation}
\begin{eqnarray}
T_{1}\partial _{\mu }\partial ^{\mu }X_{0}T_{1} &=&\partial _{\mu }\partial
^{\mu }m_{1}^{2}\,T_{1},\qquad \quad T_{2}\partial _{\mu }\partial ^{\mu
}X_{0}T_{2}=\partial _{\mu }\partial ^{\mu }m_{2}^{2}\,T_{2}\quad \quad 
\nonumber \\
T_{3}\partial _{\mu }\partial ^{\mu }X_{0}T_{3} &=&\partial _{\mu }\partial
^{\mu }m_{3}^{2}\,T_{3}\quad \quad
\end{eqnarray}

Now we can calculate for this model, the one loop effective action with non
constant background field $\phi _{cl}$ . One can choose a value for $\xi $
and proceed further. Results are known in Feynman gauge \cite{fra}, \cite{ja}%
, so we will present the results in Landau $\xi =0$ gauge. 
\begin{eqnarray}
A_{0}/(4\pi t)^{d/2} &=&\left( \delta _{i1}{\rm tr}P\,+\delta _{i2}+\delta
_{i3}\right) K(t)\,\sum_{i}{\rm tr}\left( T_{i}\right) \,e^{-m_{i}^{2}t} 
\nonumber \\
\ &=&3K_{1}(t)+K_{2}(t)+K_{2}(t)
\end{eqnarray}
\begin{eqnarray}
A_{2}/(4\pi t)^{d/2} &=&-\frac{1}{6}t^{2}\sum_{i}{\rm tr}\left( T_{i}\nabla
_{\mu }\nabla ^{\mu }X_{0}T_{i}\right) \left( \delta _{i1}{\rm tr}P\,+\delta
_{i2}+\delta _{i3}\right) \,\,K_{i}(t)-  \nonumber \\
&&\ \ \ \ \ \ \ \ \ \ -\frac{2}{3}t\,{\rm tr}\left( T_{1}\nabla _{\mu
}\nabla ^{\mu }X_{0}T_{1}\right) \,\,\widetilde{K}_{1}(t)+  \nonumber \\
&&\ \ \ \ \ \ \ \ \ \ +K(t)\,\sum_{i,j}{\rm tr}\left( T_{i}\nabla _{\mu
}X_{0}T_{i}\nabla ^{\mu }X_{0}T_{i}\right) \left( \delta _{i1}{\rm tr}%
P\,+\delta _{i2}+\delta _{i3}\right) \,\eta _{ii}(t)+  \nonumber \\
&&\ \ \ \ \ \ \ \ \ \ +\frac{2}{3}\widetilde{K}(t)\,{\rm tr}\left(
T_{1}\nabla _{\mu }X_{0}T_{1}\nabla ^{\mu }X_{0}T_{1}\right) \chi _{11}(t)+ 
\nonumber \\
&&\ \ \ \ \ \ \ \ \ \ +K(t)\,\sum_{i,j}{\rm tr}\left(
T_{i}X_{1}T_{j}X_{1}T_{i}\right) \left[ \left( \delta _{i1}+\delta
_{j1}\right) {\rm tr}P\right] \,\,\chi _{ij}(t)
\end{eqnarray}
In four dimensions we have that 
\begin{equation}
K(t)=\int_{-\infty }^{\infty }d\mu (k)\,\,e^{-k^{2}t}=\frac{1}{16\pi
^{2}t^{2}}
\end{equation}
\begin{equation}
\widetilde{K}(t)=\int_{-\infty }^{\infty }d\mu (k)\,\,\frac{1}{k^{2}}%
e^{-k^{2}t}=\frac{1}{16\pi ^{2}t}
\end{equation}
Using the values of some useful $\zeta ^{\prime }(m^{2},p,s)$ functions we
calculated in Appendix A, and the above integrals, we can find the one loop
action including the kinetic terms of the background field. We get 
\begin{eqnarray}
\zeta _{2}^{\prime }(x,0) &=&\left[ \phi _{cl}\,\phi _{cl}^{\prime \prime
}+\left( \phi _{cl}^{\prime }\right) ^{2}\right] \,\left[ 
\begin{array}{c}
\frac{7}{12\pi ^{2}}q^{2}\ln \left( \frac{m_{1}^{2}}{\mu _{R}}\right) -\frac{%
1}{64\pi ^{2}}\lambda \ln \left( \frac{m_{2}^{2}}{\mu _{R}}\right) - \\ 
-\frac{1}{192\pi ^{2}}\lambda \ln \left( \frac{m_{3}^{2}}{\mu _{R}}\right)
\end{array}
\right] +  \nonumber \\
&&  \nonumber \\
&&\ +\ \left[ \phi _{cl}^{2}\left( \phi _{cl}^{\prime }\right) ^{2}\right]
\,\left[ 
\begin{array}{c}
\frac{7}{12\pi ^{2}}q^{4}\frac{1}{m_{1}^{2}}+\frac{3}{256\pi ^{2}}\lambda
^{2}\frac{1}{m_{2}^{2}}+ \\ 
+\frac{1}{768\pi ^{2}}\lambda ^{2}\frac{1}{m_{3}^{2}}
\end{array}
\right] \ -  \nonumber \\
&&  \nonumber \\
&&\ -\left( \phi _{cl}^{\prime }\right) ^{2}\frac{q^{2}}{4\pi ^{2}}\left[ 
\begin{array}{c}
\frac{2\left[ L(m_{3}^{2})-L(m_{1}^{2})\right] }{\left(
m_{1}^{2}-m_{3}^{2}\right) ^{2}}-\frac{3}{\left( m_{1}^{2}-m_{3}^{2}\right) }%
\left[ -m_{1}^{2}+m_{1}^{2}\ln \left( \frac{m_{1}^{2}}{\mu _{R}^{2}}\right)
\right] \\ 
+\frac{3}{\left( m_{1}^{2}-m_{3}^{2}\right) }\left[ -m_{3}^{2}+m_{3}^{2}\ln
\left( \frac{m_{3}^{2}}{\mu _{R}^{2}}\right) \right]
\end{array}
\right]
\end{eqnarray}
and 
\begin{equation}
\zeta _{0}^{\prime }(x,0)=\frac{1}{16\pi ^{2}}\,\left[
3L(m_{1}^{2})+L(m_{2}^{2})+L(m_{3}^{2})\right]
\end{equation}
where 
\begin{equation}
L(m^{2})=\zeta ^{\prime }(m^{2},-2{\bf ,}0)=\frac{1}{2}m^{4}\left[ \frac{3}{2%
}-\ln \left( \frac{m^{2}}{\mu _{R}^{2}}\right) \right]
\end{equation}
Finally, the one loop corrections to the tree action are kinetic terms 
\begin{equation}
\Gamma _{2}^{(1)}=\int d^{4}x\,\left[ -{\frac{1}{2}}\zeta _{2}^{^{\prime
}}(x,0)\right]
\end{equation}
and potential terms 
\begin{equation}
\Gamma _{0}^{(1)}=\int d^{4}x\,\left[ -{\frac{1}{2}}\zeta _{0}^{^{\prime
}}(x,0)\right]
\end{equation}

In Feynman gauge the results look different, i.e. there is a $\xi $
dependence, 
\begin{eqnarray}
\zeta _{2}^{\prime }(x,0) &=&\left[ \phi _{cl}\,\phi _{cl}^{\prime \prime
}+\left( \phi _{cl}^{\prime }\right) ^{2}\right] \,\left[ 
\begin{array}{c}
\frac{1}{3\pi ^{2}}q^{2}\ln \left( \frac{m_{1}^{2}}{\mu _{R}}\right) -\frac{1%
}{64\pi ^{2}}\lambda \ln \left( \frac{m_{2}^{2}}{\mu _{R}}\right) - \\ 
-\frac{1}{192\pi ^{2}}\left( \lambda +4q^{2}\right) \ln \left( \frac{%
m_{3}^{2}}{\mu _{R}}\right)
\end{array}
\right] +  \nonumber \\
&&  \nonumber \\
&&\ \ +\ \left[ \phi _{cl}^{2}\left( \phi _{cl}^{\prime }\right) ^{2}\right]
\,\left[ 
\begin{array}{c}
\frac{1}{3\pi ^{2}}q^{4}\frac{1}{m_{1}^{2}}+\frac{3}{256\pi ^{2}}\lambda ^{2}%
\frac{1}{m_{2}^{2}}+ \\ 
+\frac{1}{768\pi ^{2}}\left( \lambda +4q^{2}\right) ^{2}\frac{1}{m_{3}^{2}}
\end{array}
\right] \ -  \nonumber \\
&&  \nonumber \\
&&\ \ -\left( \phi _{cl}^{\prime }\right) ^{2}\frac{q^{2}}{\pi ^{2}}\left[ 
\begin{array}{c}
\frac{3\left[ L(m_{3}^{2})-L(m_{1}^{2})\right] }{4\left(
m_{1}^{2}-m_{3}^{2}\right) ^{2}}-\frac{1}{\left( m_{1}^{2}-m_{3}^{2}\right) }%
\left[ -m_{1}^{2}+m_{1}^{2}\ln \left( \frac{m_{1}^{2}}{\mu _{R}^{2}}\right)
\right] \\ 
+\frac{1}{\left( m_{1}^{2}-m_{3}^{2}\right) }\left[ -m_{3}^{2}+m_{3}^{2}\ln
\left( \frac{m_{3}^{2}}{\mu _{R}^{2}}\right) \right]
\end{array}
\right]
\end{eqnarray}
while the ghost contribution is 
\begin{eqnarray}
\zeta _{2}^{^{\prime }(ghost)}(x,0) &=&\left[ \phi _{cl}\,\phi _{cl}^{\prime
\prime }+\left( \phi _{cl}^{\prime }\right) ^{2}\right] \,\left[ \frac{q^{2}%
}{48\pi ^{2}}\ln \left( \frac{m_{1}^{2}}{\mu _{R}}\right) \right]  \nonumber
\\
&&\ +\ \left[ \phi _{cl}^{2}\left( \phi _{cl}^{\prime }\right) ^{2}\right]
\,\left[ \frac{q^{4}}{48\pi ^{2}}\frac{1}{m_{1}^{2}}\right] \ 
\end{eqnarray}
and 
\begin{equation}
\zeta _{0}^{\prime }(x,0)=\frac{1}{16\pi ^{2}}\,\left[
4L(m_{1}^{2})+L(m_{2}^{2})+L(m_{3}^{2})\right]
\end{equation}
\begin{equation}
\zeta _{0}^{^{\prime }(ghost)}(x,0)=\frac{1}{16\pi ^{2}}\,L(m_{1}^{2})
\end{equation}
The one loop corrections to the tree action are the kinetic, 
\begin{equation}
\Gamma _{2}^{(1)}=\int d^{4}x\,\left[ -{\frac{1}{2}}\zeta _{2}^{^{\prime
}}(x,0)+\zeta _{2}^{^{\prime }(ghost)}(x,0)\right]  \label{kinetic}
\end{equation}
and the potential terms 
\begin{equation}
\Gamma _{0}^{(1)}=\int d^{4}x\,\left[ -{\frac{1}{2}}\zeta _{0}^{^{\prime
}}(x,0)+\zeta _{0}^{^{\prime }(ghost)}(x,0)\right]  \label{potential}
\end{equation}
The $\xi $ dependence of both the kinetic and potential terms is profound.
This dependence should cancel out for a given solution that extremises the
effective corrected action. Such a solution is the bubble one which is known
only numerically.

\section{Full electroweak model}

In order to convince the reader for the power and the range of applicability
of the described method we present the evaluation of second order derivative
terms for the full electroweak theory, around a non constant backgroung
scalar field. Recent works, performing derivative expansions, set the
Weinberg angle to zero \cite{bodeker,kripfganz,baa}, in order to simplify
the group structure of the quadratic operator.

The classical Lagrangian for the Higgs field in the electroweak model is
given by 
\begin{equation}
{\cal L}_{s}=-(D\Phi )^{\dagger }(D\Phi )+\mu ^{2}\Phi ^{\dagger }\Phi
-\lambda (\Phi ^{\dagger }\Phi )^{2}.
\end{equation}
with $\Phi $ the following $SU(2)$ doublet 
\begin{equation}
\Phi =\frac{1}{\sqrt{2}}\left( 
\begin{array}{c}
\varphi _{1}+i\varphi _{2} \\ 
\varphi _{3}+i\varphi _{4}
\end{array}
\right)
\end{equation}
Gauge covariant derivatives will be written in the form, 
\begin{equation}
D_{\mu }=\nabla _{\mu }-\frac{i}{\sqrt{2}}\,g\,A_{\mu \,a}T^{a},
\end{equation}
where 
\begin{eqnarray}
T^{a} &=&\sigma ^{a}\hbox{~~~for~~~}a=1,2,3,  \nonumber \\
T^{a} &=&t\,I\hbox{~~~for~~~}a=4.
\end{eqnarray}
The first three generators are Pauli matrices and the fourth is equal to the
unit matrix $I$ multiplied by the tangent of the Weinberg angle \cite
{salam,weinberg}. The $SU(2)$ coupling is $g$ and the $U(1)$ one $%
g^{^{\prime }}.$%
\begin{equation}
t=\frac{g^{\prime }}{g}
\end{equation}
Group indices will be raised with the metric $2\delta ^{ab}$ and lowered
with the metric $\frac{1}{2}\delta _{ab}$.

We focus on one real component $\phi $, 
\begin{equation}
\hat{\Phi}={\frac{1}{\sqrt{2}}}\pmatrix{0\cr\phi},
\end{equation}
and calculate the effective action $\Gamma [\phi ]$.

The effective action will be expanded in powers of $\hbar $, $\beta $ and $%
\nabla \phi $. First of all the effective Lagrangian including derivative
terms up to second order can be expressed as \cite{ilio}, 
\begin{equation}
{\cal L}=-\frac{1}{2}Z(\phi ,T)(\nabla \phi )^{2}-V(\phi ,T).
\end{equation}
The $\hbar $ expansion takes the form,

\begin{equation}
Z(\phi ,T)=1+Z^{(1)}(\phi ,T)  \label{ABA}
\end{equation}
\begin{equation}
V(\phi ,T)=-{\frac{1}{2}}\mu ^{2}\phi ^{2}+\frac{1}{4}\lambda \phi
^{4}+V^{(1)}(\phi ,T)  \label{ABAB}
\end{equation}
The radiative corrections to the effective potential are well known up to
this order \cite{dolan,weinberg2}. In the region of the potential where
tunnelling is important the effective Higgs mass is small and the radiative
corrections are dominated by the vector bosons and the top quark, \cite
{cari,linde}.

\subsection{Zero temperature}

We use an improved method for calculating the effective action which
includes a quadratic source \cite{bal}. This adds a term 
\begin{equation}
k=-\phi ^{-1}V^{\prime }(\phi )
\end{equation}
to the Higgs masses, making them positive, but does not change the vector
boson mass terms. At order $\hbar $, the effective action is given by 
\begin{equation}
\Gamma [\phi ]=\frac{1}{2}\log \det \Delta _{b}-\log \det \Delta _{gh}-\frac{%
1}{2}\log \det \Delta _{f}-\frac{1}{2}{\rm tr}(\Delta _{b}^{-1}k),
\end{equation}
where $\Delta _{b}$ is the fluctuation operator for the boson fields, $%
\Delta _{gh}$ for the ghosts and $\Delta _{f}$ for the fermion fields.

The gauge is fixed by the 't Hooft $R_{\xi }$ gauge fixing-functional ${\cal %
F}$. We will use 
\begin{equation}
{\cal F}_{a}=\nabla \cdot A_{a}+\frac{i}{\sqrt{2}}\xi g(\hat{\Phi}^{\dagger
}T_{a}\Phi -\Phi ^{\dagger }T_{a}\hat{\Phi}).
\end{equation}
The gauge-fixing Lagrangian is 
\begin{equation}
{\cal L}_{gf}=-\frac{1}{2\xi }{\cal F}_{a}{\cal F}^{a}
\end{equation}
The cross term of the gauge-fixing term $ig\left( \partial _{\mu }A^{a\mu
}\right) \Phi ^{\dagger }T_{a}\hat{\Phi}$ cancels a part of the cubic term $%
2ig(\partial _{\mu }\Phi ^{\dagger })T_{a}\hat{\Phi}$ $A^{a\mu }$ that
appears from the kinetic term with the covariant derivatives of the scalar
fields, after the symmetry breaking.

The quadratic term of the electroweak Lagrangian can be written as 
\begin{equation}
{\cal L}_{quad}=\frac{1}{2}\eta ^{\intercal }\Delta _{b}\,\eta -\overline{c}%
\,\Delta _{gh}\,c
\end{equation}
where 
\begin{equation}
\eta ^{\intercal }=\left( 
\begin{array}{cccccccc}
A_{1}^{\mu } & A_{2}^{\mu } & A_{3}^{\mu } & A_{4}^{\mu } & \varphi _{1} & 
\varphi _{2} & \varphi _{3} & \varphi _{4}
\end{array}
\right)
\end{equation}
and 
\begin{equation}
\Delta _{b}\equiv \Delta _{\mu \left( b\right) }^{\nu }=\left( 
\begin{array}{cc}
\Delta _{\mu \left( A\right) }^{\nu } & X_{1\mu } \\ 
X_{1}^{\nu } & \Delta _{\phi }
\end{array}
\right)
\end{equation}
The off diagonal entries are the extra kinetic terms. The fluctuations
operators are 
\begin{equation}
\Delta _{\phi }=-\nabla ^{2}+X_{\phi }
\end{equation}
\begin{equation}
\Delta _{A}\equiv \Delta _{\mu \left( A\right) }^{\nu }=-\delta _{\mu }^{\nu
}\nabla ^{2}+(1-\xi ^{-1})\nabla _{\mu }\nabla ^{\nu }+\delta _{\mu }^{\nu
}X_{A}
\end{equation}
\begin{equation}
\Delta _{gh}=-\nabla ^{2}+\xi X_{A}
\end{equation}
Let's work out to find the scalar quadratic term. The scalar quadratic part
of the gauge fixing term becomes 
\begin{eqnarray*}
&&\ \ \ \ \frac{\xi }{16}g^{2}[\;\left( 0\quad \phi \right) T_{a}\left( 
\begin{array}{c}
\varphi _{1}+i\varphi _{2} \\ 
\varphi _{3}+i\varphi _{4}
\end{array}
\right) \left( 0\quad \phi \right) T^{a}\left( 
\begin{array}{c}
\varphi _{1}+i\varphi _{2} \\ 
\varphi _{3}+i\varphi _{4}
\end{array}
\right) - \\
&&\ \ \ \ -\left( 0\quad \phi \right) T_{a}\left( 
\begin{array}{c}
\varphi _{1}+i\varphi _{2} \\ 
\varphi _{3}+i\varphi _{4}
\end{array}
\right) \left( \varphi _{1}-i\varphi _{2}\quad \varphi _{3}-i\varphi
_{4}\right) T^{a}\left( 
\begin{array}{c}
0 \\ 
\phi
\end{array}
\right) - \\
&&\ \ \ \ -\left( \varphi _{1}-i\varphi _{2}\quad \varphi _{3}-i\varphi
_{4}\right) T_{a}\left( 
\begin{array}{c}
0 \\ 
\phi
\end{array}
\right) \left( 0\quad \phi \right) T^{a}\left( 
\begin{array}{c}
\varphi _{1}+i\varphi _{2} \\ 
\varphi _{3}+i\varphi _{4}
\end{array}
\right) + \\
&&\ \ \ \ +\left( \varphi _{1}-i\varphi _{2}\quad \varphi _{3}-i\varphi
_{4}\right) T_{a}\left( 
\begin{array}{c}
0 \\ 
\phi
\end{array}
\right) \left( \varphi _{1}-i\varphi _{2}\quad \varphi _{3}-i\varphi
_{4}\right) T^{a}\left( 
\begin{array}{c}
0 \\ 
\phi
\end{array}
\right) \;] \\
\ &=&\frac{-\xi }{4}g^{2}\left[ \phi ^{2}\varphi _{1}^{2}+\phi ^{2}\varphi
_{2}^{2}+(1+t^{2})\phi ^{2}\varphi _{4}^{2}\right]
\end{eqnarray*}
Therefore 
\begin{eqnarray}
X_{0(b)} &=&\frac{\partial ^{2}V}{\partial \phi _{i}\partial \phi _{j}}-%
\frac{\xi }{4}g^{2}\left[ \phi ^{2}\varphi _{1}^{2}+\phi ^{2}\varphi
_{2}^{2}+(1+t^{2})\phi ^{2}\varphi _{4}^{2}\right]  \nonumber \\
\ &=&(3\lambda \phi ^{2}-\mu ^{2})\delta _{33}+(\lambda \phi ^{2}-\mu
^{2})(\delta _{11}+\delta _{22}+\delta _{44})-  \nonumber \\
&&\ \ \ \ -\frac{\xi }{4}g^{2}\phi ^{2}\left[ \delta _{11}+\delta
_{22}+(1+t^{2})\delta _{44}\right] +k(\delta _{11}+\delta _{22}+\delta
_{33}+\delta _{44})
\end{eqnarray}
the value of $k$ is 
\begin{equation}
k=-\lambda \phi ^{2}+\mu ^{2}
\end{equation}
Thus the eigenvalues are 
\begin{eqnarray}
m_{5}^{2} &=&m_{6}^{2}=\frac{1}{4}\xi g^{2}\phi ^{2}\;,\qquad
m_{7}^{2}=2\lambda \phi ^{2}\,,\qquad  \nonumber \\
m_{8}^{2} &=&\frac{1}{4}\xi g^{2}(1+t^{2})\phi ^{2}.
\end{eqnarray}

There are more eigenvalues of $X_{0},$ coming from $X_{A}.$ They can be
evaluated in the same way from the quadratic interaction terms with the
gauge fields $A_{\mu }.$%
\begin{equation}
m_{1}^{2}=m_{2}^{2}=\frac{1}{4}g^{2}\phi ^{2}\;,\qquad m_{3}^{2}=\frac{1}{4}%
g^{2}(1+t^{2})\phi ^{2},\qquad m_{4}^{2}=0.
\end{equation}

It is easy to see that the $T_{i}$ matrices of the expansion, Eq. \ref{X0}
are 
\begin{eqnarray}
T_{1} &=&\left( 
\begin{array}{cc}
\begin{array}{llll}
1 & 0 & 0 & 0 \\ 
0 & 0 & 0 & 0 \\ 
0 & 0 & 0 & 0 \\ 
0 & 0 & 0 & 0
\end{array}
& {\bf 0} \\ 
{\bf 0} & {\bf 0}
\end{array}
\right) \;,\qquad T_{2}=\left( 
\begin{array}{cc}
\begin{array}{llll}
0 & 0 & 0 & 0 \\ 
0 & 1 & 0 & 0 \\ 
0 & 0 & 0 & 0 \\ 
0 & 0 & 0 & 0
\end{array}
& {\bf 0} \\ 
{\bf 0} & {\bf 0}
\end{array}
\right) ,\;\qquad  \nonumber \\
T_{4} &=&T_{3}=\left( 
\begin{array}{cc}
\begin{array}{llll}
0 & 0 & 0 & 0 \\ 
0 & 0 & 0 & 0 \\ 
0 & 0 & 1 & -t \\ 
0 & 0 & -t & t^{2}
\end{array}
& {\bf 0} \\ 
{\bf 0} & {\bf 0}
\end{array}
\right) \frac{1}{1+t^{2}}\;.  \label{mat1}
\end{eqnarray}
\begin{eqnarray}
T_{5} &=&\left( 
\begin{array}{cc}
{\bf 0} & {\bf 0} \\ 
{\bf 0} & 
\begin{array}{cccc}
1 & 0 & 0 & 0 \\ 
0 & 0 & 0 & 0 \\ 
0 & 0 & 0 & 0 \\ 
0 & 0 & 0 & 0
\end{array}
\end{array}
\right) ,\qquad T_{6}=\left( 
\begin{array}{cc}
{\bf 0} & {\bf 0} \\ 
{\bf 0} & 
\begin{array}{cccc}
0 & 0 & 0 & 0 \\ 
0 & 1 & 0 & 0 \\ 
0 & 0 & 0 & 0 \\ 
0 & 0 & 0 & 0
\end{array}
\end{array}
\right) ,\   \nonumber \\
T_{7} &=&\left( 
\begin{array}{cc}
{\bf 0} & {\bf 0} \\ 
{\bf 0} & 
\begin{array}{cccc}
0 & 0 & 0 & 0 \\ 
0 & 0 & 0 & 0 \\ 
0 & 0 & 1 & 0 \\ 
0 & 0 & 0 & 0
\end{array}
\end{array}
\right) \ ,\qquad T_{8}=\left( 
\begin{array}{cc}
{\bf 0} & {\bf 0} \\ 
{\bf 0} & 
\begin{array}{cccc}
0 & 0 & 0 & 0 \\ 
0 & 0 & 0 & 0 \\ 
0 & 0 & 0 & 0 \\ 
0 & 0 & 0 & 1
\end{array}
\end{array}
\right) .  \label{mat2}
\end{eqnarray}
$\;$

The ghost quadratic part is given by the matrix $X_{gh}$ , 
\begin{equation}
X_{gh}=\xi \ \left( m_{1}^{2}T_{1}+m_{2}^{2}T_{2}+\ m_{3}^{2}T_{3}\right) .
\end{equation}
Note that $X_{gh}$ and has no spacetime indices.

The matrix $X_{1}$ is determined from the kinetic terms in the Lagrangian.
There is the following kinetic term 
\begin{eqnarray*}
&&\ \ \ -\frac{i}{\sqrt{2}}g\left( \nabla _{\mu }\phi \right) A_{\mu
a}\left( 0\quad 1\right) T^{a}\left( 
\begin{array}{c}
\varphi _{1}+i\varphi _{2} \\ 
\varphi _{3}+i\varphi _{4}
\end{array}
\right) \\
\ &=&\frac{-i}{\sqrt{2}}g\left( \nabla _{\mu }\phi \right) \left[ 
\begin{array}{c}
A_{\mu 1}\,\left( \varphi _{1}+i\varphi _{2}\right) +A_{\mu 2}\,\left(
i\varphi _{1}-\varphi _{2}\right) + \\ 
A_{\mu 3}\,\left( -\varphi _{3}-i\varphi _{4}\right) +A_{\mu 4}\left(
t\varphi _{3}+it\,\,\varphi _{4}\right)
\end{array}
\right]
\end{eqnarray*}
and also this 
\begin{eqnarray*}
&&\ \ \ \ \frac{i}{\sqrt{2}}g\left( \nabla _{\mu }\phi \right) \left(
\varphi _{1}-i\varphi _{2}\quad \varphi _{3}-i\varphi _{4}\right) A_{\mu
a}T^{a}\left( 
\begin{array}{c}
0 \\ 
1
\end{array}
\right) \\
\ &=&\frac{i}{\sqrt{2}}g\left( \nabla _{\mu }\phi \right) \left[ 
\begin{array}{c}
A_{\mu 1}\,\left( \varphi _{1}-i\varphi _{2}\right) +A_{\mu 2}\,\left(
-i\varphi _{1}-\varphi _{2}\right) + \\ 
A_{\mu 3}\,\left( -\varphi _{3}+i\varphi _{4}\right) +A_{\mu 4}\left(
t\varphi _{3}-it\,\,\varphi _{4}\right)
\end{array}
\right]
\end{eqnarray*}
Thus we get 
\begin{equation}
X_{1}\equiv X_{1\mu }=-\sqrt{2}\ g\ \,\nabla _{\mu }\phi \,\ T_{k}
\end{equation}
where 
\begin{equation}
T_{k}=\left( 
\begin{array}{cccccccc}
&  &  &  & 0 & 1 & 0 & 0 \\ 
&  &  &  & 1 & 0 & 0 & 0 \\ 
&  &  &  & 0 & 0 & 0 & -1 \\ 
&  &  &  & 0 & 0 & 0 & t \\ 
0 & 1 & 0 & 0 &  &  &  &  \\ 
1 & 0 & 0 & 0 &  &  &  &  \\ 
0 & 0 & 0 & 0 &  &  &  &  \\ 
0 & 0 & -1 & t &  &  &  & 
\end{array}
\right)
\end{equation}

We presented so far, for pedagogical reasons, how to calculate explicitly
the various matrices. In summary, we first separate off any derivative terms
already present in $X$, $X=X_{0}+X_{1}$. The eigenvalues of $X_{0}$ are the
squared particle masses $m_{i}^{2}$. Projection matrices onto the
accompanying eigenspaces are denoted by $T_{i}$.

As in the previous chapter we can define new $\Pi _{\mu }^{\nu }$ and $%
\Theta _{\mu }^{\nu }$ that satisfy similar expressions to $P_{\mu }^{\nu }$
and $Q_{\mu }^{\nu }$ . 
\begin{equation}
\Pi _{\mu }^{\lambda }\Theta _{\lambda }^{\nu }=0\;,\qquad \Pi _{\mu }^{\nu
}+\Theta _{\mu }^{\nu }=\delta _{\mu }^{\nu }\left(
T_{1}+T_{2}+T_{3}+T_{4}\right) +T_{5}+T_{6}+T_{7}+T_{8}
\end{equation}
One now can easily guess that the expansion is : 
\begin{equation}
A_{0}=(4\pi t)^{d/2}K(t)\,\sum_{i}{\rm tr}\left( T_{i}\right)
\,e^{-m_{i}^{2}t}\left[ \sum_{j=1}^{4}\delta _{ij}\left( {\rm tr}P+\xi
^{d/2}\right) \,+\sum_{j=5}^{8}\delta _{ij}\right]  \label{exp1}
\end{equation}
\begin{eqnarray}
A_{2}/(4\pi t)^{d/2} &=&-\frac{1}{6}t^{2}\sum_{i}{\rm tr}\left( T_{i}\nabla
_{\mu }\nabla ^{\mu }X_{0}T_{i}\right) \left[ \sum_{j=1}^{4}\delta
_{ij}\left( {\rm tr}P+\xi ^{-1+d/2}\right) \,+\sum_{j=5}^{8}\delta
_{ij}\right] \,\,K_{i}(t)-  \nonumber \\
&&\ \ \ \ \ -\frac{2}{3}t\sum_{i}{\rm tr}\left( T_{i}\nabla _{\mu }\nabla
^{\mu }X_{0}T_{i}\right) \sum_{j=1}^{4}\delta _{ij}\,\left[ \,\widetilde{K}%
_{i}(t)-K_{i}^{(\xi )}(t)\right]  \nonumber \\
&&\ \ \ \ \ +\,K(t)\,\sum_{i}{\rm tr}\left( T_{i}\nabla _{\mu
}X_{0}T_{i}\nabla ^{\mu }X_{0}T_{i}\right) \left[ \sum_{j=1}^{4}\delta
_{ij}\left( {\rm tr}P+\xi ^{-1+d/2}\right) \,+\sum_{j=5}^{8}\delta
_{ij}\right] \eta _{ii}(t)+  \nonumber \\
&&\ \ \ \ \ +\frac{2}{3}\left( \widetilde{K}(t)-K^{(\xi )}(t)\right)
\sum_{i,j}{\rm tr}\left( T_{i}\nabla _{\mu }X_{0}T_{j}\nabla ^{\mu
}X_{0}T_{i}\right) \sum_{k=1}^{4}\delta _{ik}\ \chi _{ij}(t)+  \nonumber \\
&&\ \ \ \ \ +K(t)\,\sum_{i,j}{\rm tr}\left( T_{i}X_{1}T_{j}X_{1}T_{i}\right)
\sum_{k=1}^{4}\left( \delta _{ik}+\delta _{jk}\right) \left[ {\rm tr}P\,+\xi
^{d/2}\right] \chi _{ij}(t)+  \nonumber \\
&&\ \ \ \ \ +\frac{2}{3}\sum_{i,j}{\rm tr}\left( T_{i}\nabla _{\mu
}X_{0}T_{j}\nabla ^{\mu }X_{0}T_{i}\right) \left[ 
\begin{array}{c}
t\,\lambda _{ij}(t)+\mu _{ij}(t)+ \\ 
+\frac{\xi }{\xi -1}\nu _{ij}(t)
\end{array}
\right] \sum_{k=1}^{4}\delta _{ik}  \label{exp2}
\end{eqnarray}
It is also plain to compute : 
\begin{equation}
{\rm tr}\left[ T_{i}\partial _{\mu }\partial ^{\mu }X_{0}T_{i}\right]
=\partial _{\mu }\partial ^{\mu }m_{i}^{2}\ ,\quad {\rm tr}\left[
T_{i}\partial _{\mu }X_{0}T_{j}\partial ^{\mu }X_{0}T_{i}\right] =\delta
_{ij}\ \left( \partial ^{\mu }m_{i}^{2}\right) ^{2}  \label{tr1}
\end{equation}
while the non zero terms of $\sum_{ij}tr[T_{i}X_{1}T_{j}X_{1}T_{i}]$ are 
\begin{eqnarray}
{\rm tr}\left[ T_{1}X_{1}T_{6}X_{1}T_{1}\right] &=&{\rm tr}\left[
T_{2}X_{1}T_{5}X_{1}T_{2}\right] ={\rm tr}\left[
T_{5}X_{1}T_{2}X_{1}T_{5}\right] ={\rm tr}\left[
T_{6}X_{1}T_{1}X_{1}T_{6}\right] =2g^{2}\left( \partial ^{\mu }\phi \right)
^{2},\quad  \nonumber \\
\quad \quad {\rm tr}\left[ T_{3}X_{1}T_{8}X_{1}T_{3}\right] &=&{\rm tr}%
\left[ T_{4}X_{1}T_{8}X_{1}T_{4}\right] ={\rm tr}\left[
T_{8}X_{1}T_{3}X_{1}T_{8}\right] =  \nonumber \\
&=&{\rm tr}\left[ T_{8}X_{1}T_{4}X_{1}T_{8}\right] =2g^{2}\left(
1+t^{2}\right) \left( \partial ^{\mu }\phi \right) ^{2}.  \label{tr2}
\end{eqnarray}
From this point on, we can calculate directly the expressions Eq. \ref{exp1}%
, Eq. \ref{exp2} and after some trivial algebra we finally compute, as in
the previous chapter, the one loop corrections from the extra kinetic and
potential terms, Eq. \ref{kinetic}, Eq.\ref{potential}, at zero temperature
and in 't Hooft gauge.

\subsection{High temperature limit}

As an example we present the results in Landau gauge ($%
m_{5}^{2}=m_{6}^{2}=m_{8}^{2}=0$ ). Masses and temperature corrections for
the boson sector are given below. The vector boson mass corrections which
include only daisy type rings \cite{cari} are

\begin{eqnarray}
m_{W^{\prime }}^{2} &=&m_{5}^{2}(T)=m_{6}^{2}(T)=\frac{1}{8}g^{2}T^{2}+\frac{%
1}{16}g^{2}(1+t^{2})T^{2}+\frac{1}{2}\lambda ^{2}T^{2}, \\
m_{W}^{2} &=&m_{1}^{2}(T)=m_{2}^{2}(T)=m_{1}^{2}+\frac{5}{6}g^{2}T^{2}, \\
m_{Z^{\prime }}^{2} &=&m_{8}^{2}(T)=\frac{1}{8}g^{2}T^{2}+\frac{1}{16}%
g^{2}(1+t^{2})T^{2}+\frac{1}{2}\lambda ^{2}T^{2}, \\
m_{H}^{2} &=&m_{7}^{2}(T)=m_{7}^{2}+\frac{1}{8}g^{2}T^{2}+\frac{1}{16}%
g^{2}(1+t^{2})T^{2}+\frac{1}{2}\lambda ^{2}T^{2}, \\
\quad m_{\gamma }^{2} &=&m_{4}^{2}(T)=\frac{1}{2}\left[ M^{2}-\sqrt{%
M^{4}-4\left( m_{1}^{2}\ \pi 1+\pi 1\ \pi 2+t^{2}m_{1}^{2}\ \pi 2\right) }%
\right] , \\
m_{Z}^{2} &=&m_{3}^{2}(T)=\frac{1}{2}\left[ M^{2}+\sqrt{M^{4}-4\left(
m_{1}^{2}\ \pi 1+\pi 1\ \pi 2+t^{2}m_{1}^{2}\ \pi 2\right) }\right] , \\
\text{with }\quad M^{2} &=&m_{1}^{2}+\pi 1+\pi 2+t^{2}m_{1}^{2}
\end{eqnarray}
where the relevant polarization tensors are given by \cite{cari}, 
\begin{equation}
\pi 1=\pi _{U(1)}=\frac{1}{6}t^{2}g^{2}T^{2},\qquad \pi 2=\pi _{SU(2)}=\frac{%
5}{6}g^{2}T^{2}
\end{equation}
It is straightforward now to calculate the $Z^{(1)}(\phi ,T)$ contribution
to the action, from the modified Eq. \ref{kinT} in the spirit of Eq. \ref
{exp1} and Eq.\ref{exp2}, using the results in Eq. \ref{tr1} and Eq. \ref
{tr2}. The fields contribute 
\begin{eqnarray}
Z^{(1)}(\phi ,T) &=&-\zeta _{2}^{\prime }=\left( \partial ^{\mu
}m_{W^{\prime }}\right) ^{2}{\frac{1}{24\pi }}\frac{T}{m_{W^{\prime }}}%
+\left( \partial ^{\mu }m_{Z^{\prime }}\right) ^{2}{\frac{1}{48\pi }}\frac{T%
}{m_{Z^{\prime }}}+\left( \partial ^{\mu }m_{H}\right) ^{2}{\frac{1}{48\pi }}%
\frac{T}{m_{H}}+  \nonumber \\
&&+\left( \partial ^{\mu }m_{W}\right) ^{2}{\frac{11}{24\pi }}\frac{T}{m_{W}}%
+\left( \partial ^{\mu }m_{\gamma }\right) ^{2}{\frac{11}{48\pi }}\frac{T}{%
m_{\gamma }}+\left( \partial ^{\mu }m_{Z}\right) ^{2}{\frac{11}{48\pi }}%
\frac{T}{m_{Z}}+  \nonumber \\
&&-g^{2}\left( \partial ^{\mu }\phi \right) ^{2}\left[ \frac{2}{%
m_{W}+m_{W^{\prime }}}+\frac{1+t^{2}}{m_{Z}+m_{Z^{\prime }}}+\frac{1+t^{2}}{%
m_{\gamma }+m_{Z^{\prime }}}\right] \frac{3T}{2\pi }\ .
\end{eqnarray}

We also present, for comparison, the kinetic term in Feynman gauge. It is
given by 
\begin{equation}
Z^{(1)}(\phi ,T)=-\zeta _{2}^{\prime (phys)}+2\zeta _{2}^{\prime (gh)}
\end{equation}
where the physical fields contribute 
\begin{eqnarray}
\zeta _{2}^{\prime (phys)} &=&\left( \partial ^{\mu }m_{W^{\prime }}\right)
^{2}{\frac{T}{48\pi }}\frac{-2}{m_{W^{\prime }}}+\left( \partial ^{\mu
}m_{Z^{\prime }}\right) ^{2}{\frac{T}{48\pi }}\frac{-1}{m_{Z^{\prime }}}%
+\left( \partial ^{\mu }m_{W}\right) ^{2}{\frac{T}{48\pi }}\frac{-8}{m_{W}}+
\nonumber \\
&&\ \ \ +\left( \partial ^{\mu }m_{H}\right) ^{2}{\frac{T}{48\pi }}\frac{-1}{%
m_{H}}+\left( \partial ^{\mu }m_{Z}\right) ^{2}{\frac{T}{48\pi }}\frac{-4}{%
m_{Z}}+  \nonumber \\
&&+g^{2}\left( \partial ^{\mu }\phi \right) ^{2}\left[ \frac{T}{\pi }\frac{1%
}{m_{W}+m_{W^{\prime }}}+\frac{T}{2\pi }\left( 1+t^{2}\right) \frac{1}{%
m_{Z}+m_{Z^{\prime }}}\right]
\end{eqnarray}
while the ghost fields give 
\begin{equation}
\zeta _{2}^{\prime (gh)}=\left( \partial ^{\mu }m_{W}\right) ^{2}{\frac{T}{%
48\pi }}\frac{-2}{m_{W}}+\left( \partial ^{\mu }m_{Z}\right) ^{2}{\frac{T}{%
48\pi }}\frac{-1}{m_{Z}}\ .
\end{equation}
Note that the ring corrections are the appropriate for this gauge choice.

Fermions do not contribute at this order. The advantage of using the ring
corrected masses is that we get finite result for small values of the scalar
field $\phi $ \cite{bodeker}, \cite{kripfganz}. The negativity of the $%
Z^{(1)}$, which is a signal of the breakdown of the pertubation theory
happens for smaller value of $\phi $ compared with the previous results, 
\cite{bodeker}, \cite{kripfganz}.

If we use the conventional approach of calculating the effective action,
without including a quadratic source, then we recover the expression in \cite
{bodeker}, using the plasma mass terms and setting $t=0$.

\section{Final remarks}

We developed this expansion method to cover fluctuation operators appearing
from a general 't Hooft gauge fixing. The superiority of these gauges is
discussed in \cite{kripfganz2}.

The aim of this work is to provide a powerful derivative expansion method
that can be used easily by other researchers in a wide range of Lagrangians.
An important feature of this derivative expansion method is that it can
handle complicated gauge groups. We applied as an example this method in the 
$SU(2)\times U(1)$ group structured model.

There is also another usefulness of the proposed method. Quantum corrections
to soliton solutions can be found applying the developed derivative
expansion technique.

As we have already pointed out the finite temperature results are only
indicative and performed for testing the method. The nonanalyticity of
Feynman amplitudes at high temperatures make the derivative expansion not
well defined \cite{das}. Widely accepted improved results still are missing.

\acknowledgments %

I would like to thank Dr. Ian Moss for enlightening discussions and for
editing the calculations.

\appendix

\section{Useful integrals and expansions.}

Let's define the following integrals in momentum space. 
\begin{equation}
K(t)=\int_{-\infty }^\infty d\mu (k)\,\,e^{-k^2t}
\end{equation}
and 
\begin{equation}
K_i(t)=K(t)\,e^{-m_i^2t}
\end{equation}

One can then prove that 
\begin{equation}
\int_{-\infty }^\infty d\mu (k)\,k_0k_0\,e^{-(k^2+m_i^2)t}=\frac 1{2t}\left(
1+\beta \frac \partial {\partial \beta }\right) K_i(t)
\end{equation}
and more general for $n=0,1,2...$ 
\begin{equation}
\int_{-\infty }^\infty d\mu (k)\,\left( k_0\right)
^{2n+2}\,e^{-(k^2+m_i^2)t}=\frac 1{2t}\left[ (2n+1)+\beta \frac \partial {%
\partial \beta }\right] \int_{-\infty }^\infty d\mu (k)\,\,\left( k_0\right)
^{2n}e^{-k^2t}  \label{anadT}
\end{equation}
where 
\begin{equation}
k^2=\overline{k}^2+k_0^2
\end{equation}
$\overline{k}$ is the spatial part of the vector $k.$ The one loop
corrections can be expressed in terms of the zeta function defined below, 
\begin{equation}
\zeta _i({\bf x,}p{\bf ,}s)\equiv \zeta _i(p{\bf ,}s)=\frac 1{\Gamma (s)}%
\int_0^\infty dt\,t^{p+s-1}\int d\mu (k)\,\,e^{-(k^2+m_i^2)t}
\end{equation}

where the $i$ subscript in $\zeta _{i}(p{\bf ,}s)$ does not represent
derivative but the index associated with the mass $m_{i}$ . The following
recursion relation helps to relate the above function with $\zeta _{i}(0{\bf %
,}s).$ 
\begin{equation}
\zeta _{i}(p+1{\bf ,}s)=-\frac{\partial }{\partial m_{i}^{2}}\zeta _{i}(p%
{\bf ,}s)  \label{ander}
\end{equation}
The value of $\zeta _{i}^{\prime }(0{\bf ,}s)$ at $s=0$ is defined by
analytic continuation. Performing a high temperature expansion $(m/T<<1)$ we
recover the well known free energy density of an ensemble of bosons or
fermions. For bosons we find 
\begin{equation}
\zeta _{i}^{\prime }(0{\bf ,}0)\simeq \frac{\pi ^{2}}{45}T^{4}-\frac{%
m_{i}^{2}}{12}T^{2}+\frac{m_{i}^{3}}{6\pi }T+\frac{m_{i}^{4}}{32\pi ^{2}}\ln
\left( \frac{\mu _{R}}{T^{2}}\right) -\frac{1}{384\pi ^{4}}\zeta _{R}(3)%
\frac{m_{i}^{6}}{T^{2}}
\end{equation}
and for fermions 
\begin{equation}
\zeta _{i}^{\prime }(0{\bf ,}0)\simeq \frac{-7}{8}\frac{\pi ^{2}}{45}T^{4}+%
\frac{m_{i}^{2}}{24}T^{2}+\frac{m_{i}^{4}}{32\pi ^{2}}\ln \left( \frac{\mu
_{R}}{T^{2}}\right) -\frac{7}{384\pi ^{4}}\zeta _{R}(3)\frac{m_{i}^{6}}{T^{2}%
}
\end{equation}

Now we will evaluate the function $\widetilde{\chi }_{ij}^{\prime }(0)$ for
bosons, defined below using the previous formulae.

\begin{eqnarray}
\widetilde{\chi }_{ij}^{\prime }(0) &=&-\frac{m^{-2}}3\frac d{ds}\left[ 
\frac 1{\Gamma (s)}\mu _R^{2s}\int dt\,t^{s-1}t\left(
e^{-m_i^2t}-e^{-m_j^2t}\right) \int d\mu (k)\,\,k^{-2}e^{-k^2t}\right] _{s=0}
\nonumber \\
\ &=&-\frac{m^{-2}}32\left[ \zeta _i^{\prime }(2{\bf ,}0)-\zeta _j^{\prime
}(2{\bf ,}0)\right]  \nonumber \\
\ &=&-\frac 23m^{-2}\left[ \frac T{8\pi }m_i^{-1}-\frac T{8\pi }%
m_j^{-1}\right]  \nonumber \\
\ &=&\frac T{12\pi }\frac 1{m_i+m_j}\frac 1{m_im_j}
\end{eqnarray}

In the same way we will calculate the function $\widetilde{\rho }%
_{i}^{\prime }(0)$ (bosons). We always try to rewrite it, using the
recursion equations, in terms of $\zeta _{i}^{\prime }(0{\bf ,}0).$

\begin{equation}
\widetilde{\rho }_{i}^{\prime }(0)=-\frac{2}{3}\frac{d}{ds}\left[ \frac{1}{%
\Gamma (s)}\mu _{R}^{2s}\int dt\,t^{s-1}t\,e^{-m_{i}^{2}t}\int d\mu
(k)\,\,k^{-2}e^{-k^{2}t}\right] _{s=0}
\end{equation}
It contains the following integral, 
\begin{equation}
I=\int_{-\infty }^{\infty }d\mu (k)\,\,k^{-2}e^{-k^{2}t}=\sum_{n=-\infty
}^{\infty }4\pi e^{-k_{0}^{2}t}\frac{1}{\beta \left( 2\pi \right) ^{3}}%
\int_{0}^{\infty }d\overline{k}\,\,\overline{k}^{2}\frac{1}{\overline{k}%
^{2}+k_{0}}e^{-\overline{k}^{2}t}\quad .
\end{equation}
The integral in the last equation can be evaluated,

\[
\int_{0}^{\infty }d\overline{k}\,\,\overline{k}^{2}\frac{1}{\overline{k}%
^{2}+k_{0}}e^{-k^{2}t}=\frac{1}{2}\sqrt{\frac{\pi }{t}}-\frac{\pi }{2}%
k_{0}e^{tk_{0}^{2}}\left[ 1-\Phi (\sqrt{t}k_{0})\right] 
\]
where 
\[
\Phi (x)=\frac{2}{\sqrt{\pi }}e^{-x^{2}}\sum_{\lambda =0}^{\infty }\frac{%
2^{\lambda }x^{2\lambda +1}}{(2\lambda +1)!!} 
\]
is the probability integral. It is also true that 
\[
I_{0}=\int_{0}^{\infty }d\overline{k}\,\,\overline{k}^{2}e^{-\overline{k}%
^{2}t}=\frac{1}{4t}\sqrt{\frac{\pi }{t}}\quad . 
\]
Combining all together we get 
\begin{eqnarray}
I &=&\sum_{n=-\infty }^{\infty }\left[ 4\pi e^{-k_{0}^{2}t}\frac{1}{\beta
\left( 2\pi \right) ^{3}}2t\,I_{0}-4\pi \frac{1}{\beta \left( 2\pi \right)
^{3}}\frac{\pi }{2}\left( \frac{2\pi n}{\beta }\right) \right]
+\sum_{n=-\infty }^{\infty }4\pi \frac{1}{\beta \left( 2\pi \right) ^{3}}%
\frac{\pi }{2}k_{0}\Phi (\sqrt{t}k_{0})\quad  \nonumber \\
&&  \nonumber \\
\ &\Rightarrow &I=2t\int_{-\infty }^{\infty }d\mu (k)\,\,e^{-(\overline{k}%
^{2}+k_{0}^{2})t}+\sum_{n=-\infty }^{\infty }\frac{4\pi }{\beta \left( 2\pi
\right) ^{3}}\frac{\pi }{2}k_{0}\frac{2}{\sqrt{\pi }}e^{-k_{0}^{2}t}\sum_{%
\lambda =0}^{\infty }\frac{2^{\lambda }t^{\left( 2\lambda +1\right)
/2}k_{0}^{2\lambda +1}}{(2\lambda +1)!!}  \nonumber \\
&&  \nonumber \\
\ &=&2t\int_{-\infty }^{\infty }d\mu (k)\,\,e^{-(\overline{k}%
^{2}+k_{0}^{2})t}++\sum_{\lambda =0}^{\infty }\frac{2^{\lambda }t^{\lambda
+1}}{(2\lambda +1)!!}4t\int_{-\infty }^{\infty }d\mu (k)\,k_{0}^{2\lambda
+2}e^{-(\overline{k}^{2}+k_{0}^{2})t}
\end{eqnarray}
We can proceed further using equation Eq. \ref{anadT}, 
\begin{equation}
I=2tK(t)+\sum_{\lambda =0}^{\infty }\frac{2^{\lambda +2}t^{\lambda +2}}{%
(2\lambda +1)!!}\frac{1}{(2t)^{\lambda +1}}\left[ \left( (2\lambda +1)+\beta 
\frac{\partial }{\partial \beta }\right) ...\left( 1+\beta \frac{\partial }{%
\partial \beta }\right) \right] K(t)\quad .
\end{equation}
Finally 
\begin{eqnarray}
\widetilde{\rho }_{i}^{\prime }(0) &=&-\frac{4}{3}\zeta _{i}^{\prime }(2{\bf %
,}0)-\frac{4}{3}\sum_{\lambda =0}^{\infty }\frac{\left( (2\lambda +1)+\beta 
\frac{\partial }{\partial \beta }\right) ...\left( 1+\beta \frac{\partial }{%
\partial \beta }\right) }{(2\lambda +1)!!}\zeta _{i}^{\prime }(2{\bf ,}0) 
\nonumber \\
&&  \nonumber \\
\ &=&-\frac{T}{6\pi }\frac{1}{m_{i}}-\frac{1}{6\pi }\frac{1}{m_{i}}%
\sum_{\lambda =0}^{\infty }\frac{\left( (2\lambda +1)+\beta \frac{\partial }{%
\partial \beta }\right) ...\left( 1+\beta \frac{\partial }{\partial \beta }%
\right) }{(2\lambda +1)!!}T  \nonumber \\
&&  \nonumber \\
\ &=&-\frac{T}{6\pi }\frac{1}{m_{i}}
\end{eqnarray}

So far we ignored the renormalisation scale in the zeta function for
simplicity. A useful definition for zero temperatures, is the following
function 
\begin{equation}
\zeta (m^{2},p{\bf ,}s)=\frac{1}{\Gamma (s)}\mu _{R}^{-2p}\int_{0}^{\infty
}dt\,t^{p+s-1}e^{-\frac{m^{2}}{\mu _{R}^{2}}t}
\end{equation}
From the definition of the Gamma functions we can find that 
\begin{eqnarray}
\zeta (m^{2},0{\bf ,}s) &=&\frac{1}{\Gamma (s)}\left( \frac{m^{2}}{\mu
_{R}^{2}}\right) ^{-s}\Gamma (s)\quad \Rightarrow  \nonumber \\
&&  \nonumber \\
\zeta ^{\prime }(m^{2},0{\bf ,}s) &=&-\left( \frac{m^{2}}{\mu _{R}^{2}}%
\right) ^{-s}\ln \left( \frac{m^{2}}{\mu _{R}^{2}}\right) \quad \Rightarrow 
\nonumber \\
&&  \nonumber \\
\zeta ^{\prime }(m^{2},0{\bf ,}0) &=&-\ln \left( \frac{m^{2}}{\mu _{R}^{2}}%
\right) \quad
\end{eqnarray}
It is obvious from the analogous recursion equation to Eq. \ref{ander}, that 
\begin{equation}
\zeta ^{\prime }(m^{2},1{\bf ,}0)=\frac{1}{m^{2}}
\end{equation}
We also can get 
\begin{eqnarray}
\zeta (m^{2},-1{\bf ,}s) &=&\frac{1}{\Gamma (s)}\mu _{R}^{2}\frac{t^{s-1}}{%
s-1}e^{-\frac{m^{2}}{\mu _{R}^{2}}t}\;\mid _{0}^{\infty }+\frac{1}{\Gamma (s)%
}m^{2}\int_{0}^{\infty }dt\frac{\,t^{s-1}}{s-1}e^{-\frac{m^{2}}{\mu _{R}^{2}}%
t}  \nonumber \\
&=&\frac{1}{s-1}m^{2}\,\zeta (m^{2},0{\bf ,}s)\qquad \Rightarrow  \nonumber
\\
&&  \nonumber \\
\zeta ^{\prime }(m^{2},-1{\bf ,}0) &=&-m^{2}+m^{2}\ln \left( \frac{m^{2}}{%
\mu _{R}^{2}}\right)
\end{eqnarray}
In the same way 
\begin{eqnarray}
\zeta (m^{2},-2{\bf ,}s) &=&\frac{1}{s-2}m^{2}\,\zeta (m^{2},-1{\bf ,}%
s)\qquad \Rightarrow  \nonumber \\
&&  \nonumber \\
\zeta ^{\prime }(m^{2},-2{\bf ,}s) &=&-\frac{1}{\left( s-2\right) ^{2}}%
m^{2}\,\zeta (m^{2},-1{\bf ,}s)+\frac{1}{s-2}m^{2}\,\zeta ^{\prime }(m^{2},-1%
{\bf ,}s)\qquad \Rightarrow  \nonumber \\
&&\qquad  \nonumber \\
\zeta ^{\prime }(m^{2},-2{\bf ,}0) &=&\frac{1}{2}m^{4}\left[ \frac{3}{2}-\ln
\left( \frac{m^{2}}{\mu _{R}^{2}}\right) \right]
\end{eqnarray}

\section{Second order term}

Here we compute the second order term in the generalized derivative
expansion. We get from Eq. \ref{iter} that 
\begin{eqnarray}
\stackrel{.}{a}_{2} &=&q+r+s+w \\
q &=&\frac{1}{2}\left( \delta ^{\rho }\Gamma ^{\sigma }-\Gamma ^{\rho
}\Gamma ^{\sigma }\right) \sum_{i,j}T_{i}X_{0,\rho \sigma }T_{j}\,e^{\left(
m_{i}^{2}-m_{j}^{2}\right) t}\;,\qquad r=\stackrel{.}{a}_{1}a_{1} \\
\ \ \ s &=&\ -\sum_{i,j}T_{i}X_{0,\rho }T_{j}\,\delta ^{\rho }a_{1}e^{\left(
m_{i}^{2}-m_{j}^{2}\right) t}\;,\qquad w=\Gamma ^{\rho
}\sum_{i,j}T_{i}X_{1,\rho }T_{j}\,e^{\left( m_{i}^{2}-m_{j}^{2}\right) t}
\end{eqnarray}

What we want now to calculate is the $A_{n}$ functions from expression Eq. 
\ref{dA}. Thus we will calculate the functions ${\rm tr}[K_{0}(k,{\bf x,}%
t)\,a_{n}(k,{\bf x},t)]$ . Since we are not interested for the \thinspace $%
A_{1}$ term we will ignore it. 
\begin{eqnarray}
{\rm tr}[K_{0}(k,{\bf x,}t)\,q] &=&-\left( \frac{1}{2}\delta ^{\rho \sigma }-%
\frac{2}{3}k^{\rho }k^{\sigma }t\right) t^{2}\left( {\rm tr}P\right) \sum_{i}%
{\rm tr}\left( T_{i}X_{0,\rho \sigma }T_{i}\right)
\,\,e^{-k^{2}t}e^{-m_{i}^{2}t}-  \nonumber \\
&&\ \ \ \ -\xi ^{-1}\left( \frac{1}{2}\delta ^{\rho \sigma }-\frac{2}{3}\xi
^{-1}k^{\rho }k^{\sigma }t\right) t^{2}\sum_{i,j}{\rm tr}\left(
T_{i}X_{0,\rho \sigma }T_{j}\right) \,\,e^{-\xi ^{-1}k^{2}t}e^{-m_{i}^{2}t}+
\nonumber \\
&&  \nonumber \\
&&\ \ \ \ +e^{-m_{i}^{2}t}\sum_{i}{\rm tr}\left( T_{i}X_{0,\rho \sigma
}T_{i}\right) \frac{1}{k^{2}}P^{,\rho \sigma }\left[ 
\begin{array}{c}
-t\left( e^{-k^{2}t}+e^{-\xi ^{-1}k^{2}t}\right) - \\ 
-\frac{2\xi }{k^{2}\left( \xi -1\right) }e^{-k^{2}t}+ \\ 
+\frac{2\xi }{k^{2}\left( \xi -1\right) }e^{-\xi ^{-1}k^{2}t}
\end{array}
\right]
\end{eqnarray}
Integrating we found 
\begin{eqnarray}
\int d\mu (k)\,{\rm tr}[K_{0}(k,{\bf x,}t)\,q] &=&-\frac{1}{6}t^{2}\left( 
{\rm tr}P\right) \sum_{i}{\rm tr}\left( T_{i}\nabla _{\rho }\nabla ^{\rho
}X_{0}T_{i}\right) \,\,K_{i}(t)-  \nonumber \\
&&-\frac{1}{6}t^{2}\xi ^{-1+d/2}\sum_{i}{\rm tr}\left( T_{i}\nabla _{\rho
}\nabla ^{\rho }X_{0}T_{i}\right) \,\,K_{i}(t)-  \nonumber \\
&&-\frac{2}{3}t\sum_{i}{\rm tr}\left( T_{i}\nabla _{\rho }\nabla ^{\rho
}X_{0}T_{i}\right) \,\,\widetilde{K}_{i}(t)+  \nonumber \\
&&+\frac{2}{3}t\sum_{i}{\rm tr}\left( T_{i}\nabla _{\rho }\nabla ^{\rho
}X_{0}T_{i}\right) \,\,K_{i}^{(\xi )}(t)  \label{q}
\end{eqnarray}
where 
\begin{equation}
K_{i}(t)=\int_{-\infty }^{\infty }d\mu
(k)\,\,e^{-k^{2}t}e^{-m_{i}^{2}t}=K(t)\,e^{-m_{i}^{2}t}
\end{equation}
\begin{equation}
\widetilde{K}_{i}(t)=\int_{-\infty }^{\infty }d\mu (k)\,\,\frac{1}{k^{2}}%
e^{-k^{2}t}e^{-m_{i}^{2}t}=\widetilde{K}(t)\,e^{-m_{i}^{2}t}
\end{equation}
and 
\begin{eqnarray}
K_{i}^{(\xi )}(t) &=&K^{(\xi )}(t)\,e^{-m_{i}^{2}t}  \nonumber \\
&&  \nonumber \\
\ &=&\int_{-\infty }^{\infty }d\mu (k)\,\left[ 
\begin{array}{c}
-k^{2}\,e^{-\xi ^{-1}k^{2}t}+ \\ 
\frac{1}{k^{4}}\frac{4}{\left( 1-\xi ^{-1}\right) t}\sinh \left( \frac{k^{2}%
}{2}(1-\xi ^{-1})t\right) e^{-\frac{k^{2}}{2}(1+\xi ^{-1})t}
\end{array}
\right] e^{-m_{i}^{2}t}
\end{eqnarray}
From the above definitions it is easy to check that 
\begin{equation}
\text{for }\xi =0\qquad K_{i}^{(\xi )}(t)=0
\end{equation}
\begin{equation}
\text{for }\xi =1\qquad K_{i}^{(\xi )}(t)=\widetilde{K}_{i}(t)
\end{equation}
From the second term of $\stackrel{.}{a}_{2}$ get 
\begin{eqnarray}
&&\ \ {\rm tr}[K_{0}(k,{\bf x,}t)\,r]  \nonumber \\
&=&-4k^{\rho }k^{\sigma }\left[ 
\begin{array}{c}
{\rm tr}\left( P\right) e^{-k^{2}t}+ \\ 
+\xi ^{-2}e^{-\xi ^{-1}k^{2}t}
\end{array}
\right] S_{i,j}^{0}\,\left[ 
\begin{array}{c}
\frac{t^{3}}{3}+\frac{t^{2}}{2m^{2}}- \\ 
-\frac{1}{m^{2}}f(m^{2})
\end{array}
\right] \,\frac{-1}{m^{2}}e^{-m_{i}^{2}t}+  \nonumber \\
&&  \nonumber \\
&&+Z_{i,j}\,e^{-k^{2}t}\left[ 
\begin{array}{c}
t-g(m^{2})-g(k^{2}(1-\xi ^{-1}))+ \\ 
+g(m^{2}+k^{2}(1-\xi ^{-1}))
\end{array}
\right] \frac{-1}{m^{2}}e^{-m_{i}^{2}t}-  \nonumber \\
&&  \nonumber \\
&&-Z_{i,j}e^{-k^{2}t}\left[ 
\begin{array}{c}
-t-g(m^{2})+g(-k^{2}(1-\xi ^{-1}))+ \\ 
+g(m^{2}+k^{2}(1-\xi ^{-1}))
\end{array}
\right] \frac{-1}{m^{2}+k^{2}(1-\xi ^{-1})}e^{-m_{i}^{2}t}-  \nonumber \\
&&  \nonumber \\
&&-Z_{i,j}e^{-\xi ^{-1}k^{2}t}\left[ 
\begin{array}{c}
-t+g(m^{2})+g(-k^{2}(1-\xi ^{-1}))- \\ 
-g(m^{2}-k^{2}(1-\xi ^{-1}))
\end{array}
\right] \frac{-1}{m^{2}}e^{-m_{i}^{2}t}+  \nonumber \\
&&  \nonumber \\
&&+Z_{i,j}e^{-\xi ^{-1}k^{2}t}\left[ 
\begin{array}{c}
t+g(m^{2})-g(k^{2}(1-\xi ^{-1}))- \\ 
-g(m^{2}-k^{2}(1-\xi ^{-1}))
\end{array}
\right] \frac{1}{-m^{2}+k^{2}(1-\xi ^{-1})}e^{-m_{i}^{2}t}+  \nonumber \\
&&  \nonumber \\
&&+\left[ 
\begin{array}{c}
{\rm tr}\left( P\right) e^{-k^{2}t}+ \\ 
+e^{-\xi ^{-1}k^{2}t}
\end{array}
\right] S_{i,j}^{1}\,\,\left[ t-g(m^{2})\right] \,\frac{-1}{m^{2}}%
e^{-m_{i}^{2}t}+\text{odd terms}
\end{eqnarray}
where 
\begin{equation}
Z_{i,j}=k^{-2}P^{,\rho \sigma }S_{i,j}^{0}=k^{-2}P^{,\rho \sigma }\sum_{i,j}%
{\rm tr}\left( T_{i}X_{0,\rho }T_{j}X_{0,\sigma }T_{i}\right)
\end{equation}
\begin{equation}
S_{i,j}^{1}=\sum_{i,j}{\rm tr}\left( T_{i}X_{1}T_{j}X_{1}T_{i}\right)
\end{equation}
The odd terms, proportional to $k^{\rho }$ do not survive after the
integration : 
\begin{eqnarray}
&&\ \ \int d\mu (k)\,{\rm tr}[K_{0}(k,{\bf x,}t)\,r]  \nonumber \\
&=&\frac{2}{t}\left[ {\rm tr}\left( P\right) +\xi ^{-1+d/2}\right]
S_{i,j}^{0}\frac{1}{m^{2}}K_{i}(t)\left[ 
\begin{array}{c}
\frac{t^{3}}{3}+\frac{t^{2}}{2m^{2}}- \\ 
-\frac{1}{m^{2}}f(m^{2})
\end{array}
\right] +  \nonumber \\
&&  \nonumber \\
&&+\frac{2}{3}S_{i,j}^{0}\frac{1}{m^{2}}\widetilde{K}_{i}(t)\,g(m^{2})\,%
\left( 1+\xi ^{-1+d/2}\right) +  \nonumber \\
&&  \nonumber \\
&&+\frac{2}{3}S_{i,j}^{0}\frac{1}{m^{2}}\widetilde{K}_{i}(t)\,g(m^{2})\,%
\left[ 
\begin{array}{c}
\Lambda _{i}(m^{2},t)-\Lambda _{i}(-m^{2},\xi ^{-1}t)+ \\ 
+e^{m^{2}t}\left[ \Lambda _{i}(-m^{2},t)-\Lambda _{i}(m^{2},\xi
^{-1}t)\right]
\end{array}
\right] +  \nonumber \\
&&  \nonumber \\
&&+\frac{2}{3}S_{i,j}^{0}\left( g(m^{2})+t\right) \,\left[ -\Lambda
_{i}(m^{2},t)+\Lambda _{i}(-m^{2},\xi ^{-1}t)\right] +  \nonumber \\
&&  \nonumber \\
&&+\frac{2}{3}S_{i,j}^{0}\frac{\xi }{\xi -1}\,\left[ 
\begin{array}{c}
\widetilde{\Lambda }_{i}(m^{2},t)+\widetilde{\Lambda }_{i}(-m^{2},\xi
^{-1}t)- \\ 
-\widetilde{\Lambda }_{i}(m^{2},\left( 2-\xi ^{-1}\right) t)-\widetilde{%
\Lambda }_{i}(-m^{2},\left( 2\xi ^{-1}-1\right) t)
\end{array}
\right] +  \nonumber \\
&&  \nonumber \\
&&+\frac{2}{3}S_{i,j}^{0}\,\left[ 
\begin{array}{c}
-N_{i}(m^{2},t)-N_{i}(-m^{2},\xi ^{-1}t)+ \\ 
+e^{m^{2}t}\left[ N_{i}(m^{2},\xi ^{-1}t)+N_{i}(-m^{2},t)\right]
\end{array}
\right] -  \nonumber \\
&&  \nonumber \\
&&-\frac{2}{3}t\,S_{i,j}^{0}\widetilde{K}_{i}(t)\frac{1}{m^{2}}+\frac{2}{3}%
t\,S_{i,j}^{0}K_{i}^{(\xi )}(t)\frac{1}{m^{2}}-  \nonumber \\
&&\ -\left[ {\rm tr}\left( P\right) +\xi ^{d/2}\right] S_{i,j}^{1}\frac{1}{%
m^{2}}K_{i}(t)\left( t-g(m^{2})\right)  \label{r}
\end{eqnarray}
where 
\begin{equation}
\Lambda _{i}(m^{2},t)=\int_{-\infty }^{\infty }d\mu (k)\,\,k^{-2}\frac{1}{%
m^{2}+k^{2}(1-\xi ^{-1})}e^{-k^{2}t}e^{-m_{i}^{2}t}
\end{equation}
\begin{equation}
\widetilde{\Lambda }_{i}(m^{2},t)=\int_{-\infty }^{\infty }d\mu (k)\,\,k^{-4}%
\frac{1}{m^{2}+k^{2}(1-\xi ^{-1})}e^{-k^{2}t}e^{-m_{i}^{2}t}
\end{equation}
\begin{equation}
N_{i}(m^{2},t)=\int_{-\infty }^{\infty }d\mu (k)\,\,k^{-2}\frac{1}{\left[
m^{2}+k^{2}(1-\xi ^{-1})\right] ^{2}}e^{-k^{2}t}e^{-m_{i}^{2}t}
\end{equation}
We focus now on the third term of $\stackrel{.}{a}_{2}$, 
\begin{eqnarray}
&&\ \ \ {\rm tr}[K_{0}(k,{\bf x,}t)\,s]  \nonumber \\
\ &=&2\left[ 
\begin{array}{c}
{\rm tr}\left( P\right) e^{-k^{2}t}+ \\ 
+\xi ^{-1}e^{-\xi ^{-1}k^{2}t}
\end{array}
\right] S_{i,j}^{0}\,\left[ 
\begin{array}{c}
\frac{t^{2}}{2}+\frac{t}{m^{2}}- \\ 
-\frac{1}{m^{2}}g(m^{2})
\end{array}
\right] \,\frac{-1}{m^{2}}e^{-m_{i}^{2}t}+  \nonumber \\
&&  \nonumber \\
&&+\left( -e^{-k^{2}t}+e^{-\xi ^{-1}k^{2}t}\right) Z_{i,j}\,\left[
-g(m^{2})+g(k^{2}(1-\xi ^{-1}))\right] \frac{e^{-m_{i}^{2}t}}{%
-m^{2}+k^{2}(1-\xi ^{-1})}+  \nonumber \\
&&  \nonumber \\
&&+\left( -e^{-k^{2}t}+e^{-\xi ^{-1}k^{2}t}\right) Z_{i,j}\,\left[
-g(m^{2})+g(-k^{2}(1-\xi ^{-1}))\right] \frac{e^{-m_{i}^{2}t}}{%
m^{2}+k^{2}(1-\xi ^{-1})}+  \nonumber \\
&&+\text{odd terms}
\end{eqnarray}
and after the integration 
\begin{eqnarray}
&&\ \ \int d\mu (k)\,{\rm tr}[K_{0}(k,{\bf x,}t)\,s]  \nonumber \\
&=&-2\left[ {\rm tr}\left( P\right) +\xi ^{-1+d/2}\right] S_{i,j}^{0}\frac{1%
}{m^{2}}K_{i}(t)\left[ 
\begin{array}{c}
\frac{t^{2}}{2}+\frac{t}{m^{2}}- \\ 
-\frac{1}{m^{2}}g(m^{2})
\end{array}
\right] +  \nonumber \\
&&  \nonumber \\
&&+\frac{2}{3}S_{i,j}^{0}\frac{\xi }{\xi -1}\,\left[ 
\begin{array}{c}
-2\widetilde{\Lambda }_{i}(-m^{2},\xi ^{-1}t)+\widetilde{\Lambda }%
_{i}(-m^{2},t)+\widetilde{\Lambda }_{i}(-m^{2},\left( 2\xi ^{-1}-1\right) t)+
\\ 
+\widetilde{\Lambda }_{i}(m^{2},\left( 2-\xi ^{-1}\right) t)-2\widetilde{%
\Lambda }_{i}(m^{2},t)+\widetilde{\Lambda }_{i}(m^{2},\xi ^{-1}t)
\end{array}
\right] +  \nonumber \\
&&  \nonumber \\
&&+\frac{2}{3}S_{i,j}^{0}\,g(m^{2})\,\left[ 
\begin{array}{c}
\Lambda _{i}(-m^{2},t)-\Lambda _{i}(-m^{2},\xi ^{-1}t) \\ 
+\Lambda _{i}(m^{2},t)+\Lambda _{i}(m^{2},\xi ^{-1}t)
\end{array}
\right]  \label{s}
\end{eqnarray}
The contribution of the fourth term $w$ , of $\stackrel{.}{a}_{2}$ is zero
because the trace gives a function proportional to $k^{\rho }$ .

Finally from Eq. \ref{q}, Eq. \ref{r}, Eq. \ref{s} after considerable
cancellations, we find through the expression Eq. \ref{dA} that in 't Hooft
gauge the first terms of the derivative expansion are given by expression
Eq. \ref{onet}, Eq. \ref{twot}.

\end{document}